\documentclass[twocolumn]{aastex61}

\usepackage{rotating}
\usepackage{amssymb}
\usepackage{amsfonts}
\usepackage{amsmath}
\usepackage[varg]{txfonts}
\usepackage{natbib}
\usepackage{color}
\usepackage{comment}
\usepackage{graphicx,rotating}

\usepackage{longtable}
\usepackage{threeparttablex}
\usepackage{lineno}
\newcommand{\RNum}[1]{\uppercase\expandafter{\romannumeral #1\relax}}
\newcommand{\lamost}[1]{\textsc{F\small{AMOST}}}

\newcommand\aastex{AAS\TeX}

\received{\today}
\revised{---}
\accepted{---}
\submitjournal{AAS Journal}

\shorttitle{\aastex\ Zong et al. (2020) The LK-MRS survey}
\shortauthors{Zong et al. (2020)}

\begin{document}

\title{
%Phase \RNum{2} of the LAMOST-{\sl Kepler}/{\sl K}2 survey. \RNum{1}. Time series medium-resolution spectroscopic observations
Phase \RNum{2} of the LAMOST-{\sl Kepler}/{\sl K}2 survey. \RNum{1}. Time series of medium-resolution spectroscopic observations
}

\correspondingauthor{Jian-Ning Fu}
\email{jnfu@bnu.edu.cn}

\author{Weikai Zong}
\affil{Department of Astronomy, Beijing Normal University, Beijing~100875, P.~R.~China}

\author{Jian-Ning Fu}
\affil{Department of Astronomy, Beijing Normal University, Beijing~100875, P.~R.~China}

\author{Peter De Cat}
\affil{Royal Observatory of Belgium, Ringlaan 3, B-1180 Brussel, Belgium}

\author{Jiaxin Wang}
\affil{Department of Astronomy, Beijing Normal University, Beijing~100875, P.~R.~China}

\author{Jianrong Shi}
\affil{Key Lab for Optical Astronomy, National Astronomical Observatories, Chinese Academy of Sciences, Beijing 100012, P.~R.~China}

\author{Ali Luo}
\affil{Key Lab for Optical Astronomy, National Astronomical Observatories, Chinese Academy of Sciences, Beijing 100012, P.~R.~China}

\author{Haotong Zhang}
\affil{Key Lab for Optical Astronomy, National Astronomical Observatories, Chinese Academy of Sciences, Beijing 100012, P.~R.~China}

\author{A. Frasca}
\affil{INAF -- Osservatorio Astrofisico di Catania, Via S. Sofia 78, I-95123 Catania, Italy}

\author{J. Molenda- \.Zakowicz}
\affil{Astronomical Institute of the University of Wroc\l{}aw, ul. Kopernika 11, 51-622 Wroc\l{}aw, Poland}

\author{R. O. Gray}
\affil{Department of Physics and Astronomy, Appalachian State University, Boone, NC 28608, USA}

\author{C. J. Corbally}
\affil{Vatican Observatory Research Group, Steward Observatory, Tucson, AZ 85721-0065, USA}

\author{G. Catanzaro}
\affil{INAF -- Osservatorio Astrofisico di Catania, Via S. Sofia 78, I-95123 Catania, Italy}

\author{Tianqi Cang}
\affil{IRAP, Universit\'e de Toulouse, CNRS, UPS, CNES, 14 avenue Edouard Belin, F-31400, Toulouse, France}

\author{Jiangtao Wang}
\affil{Department of Astronomy, Beijing Normal University, Beijing~100875, P.~R.~China}

\author{Jianjun Chen}
\affil{Key Lab for Optical Astronomy, National Astronomical Observatories, Chinese Academy of Sciences, Beijing 100012, P.~R.~China}

\author{Yonghui Hou}
\affil{Nanjing Institute of Astronomical Optics \& Technology, National Astronomical Observatories, Chinese Academy of Sciences, Nanjing~210042, P.~R.~China}

\author{Jiaming Liu}
\affil{Key Lab for Optical Astronomy, National Astronomical Observatories, Chinese Academy of Sciences, Beijing 100012, P.~R.~China}

\author{Hubiao Niu}
\affil{Department of Astronomy, Beijing Normal University, Beijing~100875, P.~R.~China}
\affil{Xinjiang Astronomical Observatory, Chinese Academy of Sciences, Urumqi, Xinjiang~830011, P.~R.~China}

\author{Yang Pan}
\affil{Department of Astronomy, Beijing Normal University, Beijing~100875, P.~R.~China}

\author{Hao Tian}
\affil{Key Lab for Optical Astronomy, National Astronomical Observatories, Chinese Academy of Sciences, Beijing 100012, P.~R.~China}

\author{Hongliang Yan}
\affil{Key Lab for Optical Astronomy, National Astronomical Observatories, Chinese Academy of Sciences, Beijing 100012, P.~R.~China}

\author{Yong Zhang}
\affil{Nanjing Institute of Astronomical Optics \& Technology, National Astronomical Observatories, Chinese Academy of Sciences, Nanjing~210042, P.~R.~China}

\author{Heng Zuo}
\affil{Nanjing Institute of Astronomical Optics \& Technology, National Astronomical Observatories, Chinese Academy of Sciences, Nanjing~210042, P.~R.~China}

%\author{LAMOST-{\sl Kepler} collaboration}


%%\Large
\begin{abstract}
Phase \RNum{2} of the LAMOST-{\sl Kepler/K}2 survey (LK-MRS), initiated in 2018, aims at collecting medium-resolution spectra ($R\sim7,500$; hereafter MRS) for more than $50,000$ stars with multiple visits ($\sim60$ epochs) over a period of 5 years (2018 September to 2023 June). 
We selected 20 footprints distributed across the {\sl Kepler} field and six {\sl K}2 campaigns, with each plate containing a number of stars ranging from $\sim2,000$ to $\sim 3,000$. 
During the first year of observations, the LK-MRS has already visited 13 plates 223 times over 40 individual nights, and collected $\sim280,000$ and $\sim369,000$ high-quality spectra in the blue and red wavelength range, respectively. 
The atmospheric parameters and radial velocities for $\sim259,000$ spectra of $21,053$ targets were successfully calculated by the LASP pipeline.
The internal uncertainties for the effective temperature, surface gravity, metallicity, and radial velocity are found to be $100$\,K, $0.15$\,dex, $0.09$\,dex, and $1.00$\,km\,s$^{-1}$, respectively, when derived from a medium-resolution LAMOST spectrum with a signal-to-noise ratio in the $g$-band (S/N) of 10. 
All the uncertainties decrease as S/N increases, but they stabilize for ${\rm S/N}>100$. 
We found $14,997$, $20,091$, and $1,514$ stars in common with the targets from the LAMOST low-resolution survey (LRS), GAIA and APOGEE, respectively, corresponding to a fraction of $\sim70\%$, $\sim95\%$ and $\sim7.2\%$. 
In general, the parameters derived from LK-MRS spectra are consistent with those obtained from the LRS and APOGEE spectra, but the scatter increases as the surface gravity decreases when comparing with the measurements from APOGEE.
A large discrepancy is found with the GAIA values of the effective temperature.
The comparisons of radial velocities of LK-MRS to GAIA and LK-MRS to APOGEE nearly follow an Gaussian distribution with a mean $\mu\sim1.10$ and $0.73$\,km\,s$^{-1}$, respectively. 
We expect that the results from the LK-MRS spectra will shed new light on binary stars, asteroseismology, stellar activity, and other research fields.
\end{abstract}


\section{Introduction}
Planetary science and stellar physics have benefited from large photometric \citep[see, e.g.,][]{kepler2010,k22014,tess2014}, spectroscopic \citep[see, e.g.,][]{sdss2015,lamost2015} and astrometric \citep[see, e.g.,][]{1997A&A...323L..49P,gaia2016a,gaia2018} surveys. For 
instance, in the realm of pulsating star physics, the analysis of high quality photometric data, in particular from space platforms, can
yield a set of frequencies resolved down to a precision of a few nHz \citep[see, e.g.,][]{2015MNRAS.454.1792K,2017MNRAS.465.1057K,zong2016a,zong2016b,zong2018a}. 
Combined with atmospheric parameters determined from spectroscopy, the technique of asteroseismology may be used with the results of
precision photometry to study the interior of pulsating stars with unprecedented precision \citep[see, e.g.,][]{2018Natur.554...73G}. 
While those seismic results can be used to calibrate some key physical processes, such as the rates of nuclear reactions during stellar evolution \citep[see, e.g.,][]{2016ApJ...823...46F}, the reliability of those same seismic solutions can be tested by comparing the asteroseismic distances with those determined from astrometry \citep[see, e.g.,][]{charpinet2019}. 
As for planetary science, large photometric surveys allow a statistical analysis of planetary properties as well as how those properties relate to the properties of their host stars
\citep[see, e.g.,][]{2013ApJS..204...24B,2018AJ....156..221N}. 
%%%AF:
However, even if the space-based photometry has an unprecedented high precision, the photometric solution for the planetary properties 
may still have large uncertainties if some of the fundamental parameters of the host star are poorly known \citep[see, e.g.,][]{2014ApJS..211....2H}. 
Combining the accurate properties of host stars derived from spectroscopy with precise parallaxes and distances, will greatly reduce the errors propagating to the characterization of planets, revealing, for example, clear relationships between 
planetary and stellar properties
\citep[see., e.g.,][]{2019ApJ...875...29M,2018MNRAS.480.2206O}.

Spaceborne precision photometry began with a trio of missions:
MOST \citep[Microvariability and Oscillations of STars,][]{2003PASP..115.1023W}, CoRoT \citep[Convection, Rotation and Transit experiment,][]{2009A&A...506..411A}, and {\sl Kepler}.
The {\sl NASA} mission {\sl Kepler}, launched in 2009 March and operational until 2019 May, delivered photometric data with unprecedented high quality for more than $780,000$ targets \citep{2018arXiv181012554B}. 
{\sl Kepler} was designed to detect Earth-sized planets around Solar-like stars within a 105 deg$^2$ field in the region between the constellations of Cygnus and Lyrae \citep{kepler2010}. 
Its high-quality photometry is also a goldmine for the field of asteroseismology \citep[see, e.g.,][]{gilliand10} as well as for many other science cases \citep[see, e.g., eclipsing binaries in][]{2011AJ....141...83P}. 
However, in 2013 May, the spacecraft lost the second of its four reaction wheels on board, ending the main mission.  A follow-on mission (the {\sl K}2 mission), with precision pointing provided by the two remaining reaction wheels and radiation pressure from the sun was designed to point towards
20 fields along the ecliptic plane, with each campaign (C0, C1,..., C19) having a duration of $\sim80$ days, in period from 2014 to 2018 \citep{k22014}. The {\sl K}2 mission opened the door to more scientific topics compared to the original {\sl Kepler} mission, covering, e.g., more pulsating white dwarfs \citep{2017ApJS..232...23H}, the first transit event around a white dwarf \citep{2015Natur.526..546V}, microlensing events \citep[see, e.g.,][]{2016PASP..128l4401H}, and accreting young stellar objects \citep[e.g.][]{2018AJ156..71}.

It should be kept in mind that the {\sl Kepler Input Catalog} \citep[KIC;][]{2011AJ....142..112B} provides rather low-precision atmospheric parameters for objects in the {\sl Kepler} field of view and that the {\sl Ecliptic Plane Input Catalog} \citep[EPIC;][]{2016ApJS..224....2H} only contains basic properties of input targets for the {\sl K}2 campaigns, though they were revisited and revised by \citet{2014ApJS..211....2H,2016ApJS..224....2H}. 
Therefore, a number of follow-up spectroscopic observations have been performed to improve the precision of the atmospheric parameters and/or the radial velocities for the targets with {\sl Kepler/K}2 photometry \citep[see, e.g.,][]{2010AN....331..993U,2012A&A...543A.160T,2015MNRAS.450.2764N,2018ApJ...861..149F,2019MNRAS.484..451H}. 
A homogeneous study was performed specifically on the 1305 stars hosting 2075 planets with the Keck high-resolution spectrograph HIRES \citep[the California-$Kepler$ Survey;][]{2017AJ....154..107P}.
Due to the large number of targets, it is necessary for the ground-based telescopes to employ multiple fibers with the aim of observing as many of the {\sl Kepler/K}2 targets as possible in an efficient way. 
Such endeavors have been carried out on the Sloan Digital Sky Survey ({\sl SDSS}), the Large Sky Area Multi-Object Fibre Spectroscopic Telescope (LAMOST), and the Anglo-Australian telescope (AAT) in the framework of the APOKASC survey \citep{2017ApJS..233...23S,2018ApJS..239...32P}, the LAMOST-{\sl Kepler} project \citep{2015ApJS..220...19D,zong2018b}, and the {\sl K}2-HERMES survey \citep{2018AJ....155...84W}, respectively.

In 2011, the LAMOST-{\sl Kepler} (hereafter LK) project was initiated with the aim to use LAMOST as a follow-up telescope to collect spectroscopic observations for as many objects in the {\sl Kepler} field as possible \citep[see details in][]{2015ApJS..220...19D}. 
From the first five-year regular survey (2012-2017), the LK project obtained $227,870$ low-resolution spectra of $156,390$ stars, including a fraction of $\sim40\,\%$ of the {\sl Kepler} targets \citep[][hereafter Z18b]{zong2018b}. 
Those spectra were analyzed through three different pipelines: 
(i) the LAMOST stellar parameter pipeline \citep[LASP;][]{2011A&A...525A..71W,2014IAUS..306..340W,lamost2015}; 
(ii) an updated version of the code ROTFIT \citep{2003A&A...405..149F,2006A&A...454..301F,2016A&A...594A..39F}; 
and (iii) the code MKCLASS for an automatic spectral classification \citep{2014AJ....147...80G,2016AJ....151...13G}. 
The atmospheric parameters derived from the high-quality spectra (S/N$\sim50$) of objects that have been visited multiple times have a precision of $\sim95$~K for the effective temperature $T_\mathrm{eff}$, $\sim0.11$~dex for the surface gravity $\log g$, and $\sim0.09$~dex for the metallicity [Fe/H] \citep{2016ApJS..225...28R}. 
This large library of spectra with derived quantities has received much attention from various research fields, both for statistical studies \citep[see, e.g.,][]{2016NatCo...711058K,2015MNRAS.453.1095B,2018PNAS..115..266D,2016AJ....152..187M} and for the study of individual stars \citep[see, e.g.,][]{2014A&A...564A..27D,2016ApJ...827L..17M,2018MNRAS.477.2020C}.  
From 2015 onwards, observations have been collected for the 
LAMOST-{\sl K}2 project (hereafter LK2; J.~Wang, et al. 2020, in prep.). 
The LK2 project is similar to the LK project but the footprints 
point towards the {\sl K}2 campaigns with declinations higher than $-10^\circ$. 
So far, $\sim160,000$ spectra for $\sim85,000$ different {\sl K}2 targets have been collected in the framework of the LK2 project.

The first phase of the regular survey of LAMOST ended in 2017 June. From that September, LAMOST was equipped and tested with medium-resolution spectrographs ($R\sim7500$), each one with a blue and a red arm, which range from $495$ to $535$\,nm and from $630$ to $680$\,nm, respectively (\citealt{2019RAA....19...75L}; hereafter L19). 
A first analysis revealed that the precision of the radial velocity (RV) is close to $1$\,km\,s$^{-1}$. This estimation was obtained by L19, who analyzed {the RV scatter of stars with standard deviation less than 0.5\,km\,s$^{-1}$ from} $\sim1900$ targets with multiple MRS spectra. This confirms our expectations, based on the higher resolution,
that the precision of RV values derived from MRS spectra is $\sim3-5$ times better than for those obtained from the LRS spectra \citep{lamost2015}. 
The MRS survey was approved to be performed, along with the existing low-resolution one, in the second phase of the regular survey of LAMOST from 2018 September to 2023 June.

Among several independent programs in that setup, we initiated the LK-MRS survey with the aim to obtain time-series of medium-resolution spectra for a selection of 20 footprints. 
This paper is the first of a series dedicated to the description and analysis of the spectra obtained within the LK-MRS survey. 
Here we focus on the data collected between 2018 May and 2019 June. 
The structure of this paper is organized as follows. 
In Section~2, we describe the LK-MRS survey, including details on the observations and the quality of the spectra obtained in the first year.
The description of the database of atmospheric parameters ($T_\mathrm{eff}$, $\log g$, and [Fe/H]) and RVs derived from the MRS spectra is given in Section~3 while the associated evaluation of the internal uncertainties for those four quantities and their comparison with other large surveys are presented in Section~4.
Section~5 includes the discussion of the prospects for several scientific aspects of the LK-MRS survey.
We end with a brief summary in Section~6. 

% ------------------------------------------------------------
\section{Phase \RNum{2} of the LAMOST-{\sl Kepler/K}2 Survey}

\subsection{Project description}
% ------------------------------------------------------------------
\begin{figure}
\centering
%\epsscale{.80}
\includegraphics[width=8cm]{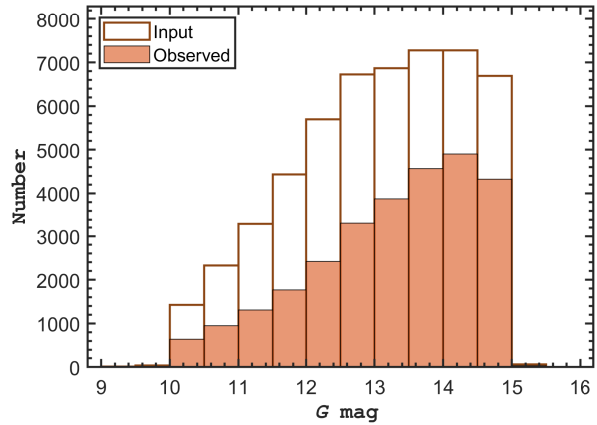}
\caption{
$G$-band magnitude distribution with a bin width of 0.5\,mag for all stars in the LK-MRS input catalog (white) and those for which at least one medium-resolution LAMOST spectrum with S/N$>10$ is already available (orange).  
We note that a few targets fainter than 15th magnitude have been observed.
}
\label{fig1}
\end{figure}
% ====================================================================
The prime goal of the LK-MRS project is to provide precise atmospheric parameters ($T_\mathrm{eff}$, $\log g$, and [Fe/H]) and RVs for stars distributed in the fields of the {\sl Kepler} and {\sl K}2 campaigns with LAMOST\footnote{ The Large Sky Area Multi-Object Fiber Spectroscopic Telescope \citep[also called Gou Shoujing Telescope;][]{1996ApOpt..35.5155W,1998SPIE.3352..839X} which is located at the Xinglong Observatory, China. It has a field of view in diameter of 5 degrees and is equipped with 4000 fibers at the focus.}. 
From 2018 September to 2023 June, it has been approved for LAMOST to collect LRS and MRS spectra in parallel. 
Whereas the LRS spectra are taken in the dark nights of each lunar month (from night 23 to night 6 of the next lunar month), the other nights are reserved for MRS observations. 
Within the MRS working group, a distinction is made between time domain (TD) observations and non-TD (NT) ones \citep{2020arXiv200507210L}. 
The LK-MRS survey is one of four TD projects, for which about $120$\,hrs of LAMOST time is allocated annually. 
This corresponds to the time needed to observe 60 plates. 
Indeed, each plate typically requires $\sim2$\,hrs of LAMOST time, consisting of $\sim30$\,min of overhead time (pointing) and $\sim90$\,min of observing time (sufficient for 4 exposures of $20$\,min and $3$\,min read-out time for each exposure). Based on this time allocation, we designed a strategy to observe 20 footprints at about 60 epochs each in a time span of 5 years.
The selection of the footprints for the LK-MRS survey depends on three conditions. 
(1) LAMOST can only observe the field from two hours before to two hours after its meridian passage.
(2) The declination of the field must be higher than $-10$ degrees.
(3) The distribution of the plates needs to be as homogeneous as possible in right ascension,
which will reduce the conflict with observations for other LAMOST projects.
With those criteria in mind, we selected 4 footprints in the {\sl Kepler} field and a total of 16 footprints in the {\sl K}2 campaigns (C4, C5, C8, C13, C14, and C16). 
The central position of each footprint is determined by the coordinate of its central star, which must be brighter than 8th magnitude in the $V$-band. 
This latter requirement is also valid for LRS projects such as the LK project \citep{2015ApJS..220...19D,zong2018b}. Each footprint contains flux standard stars, targets of scientific interest and fibres for sky background measurements.
 
For the prioritisation of the targets within the selected footprints, the highest priority is given to stars with {\sl Kepler/K}2 photometry. 
The objects in the four most central footprints of the LK project were chosen as the targets in the {\sl Kepler} field for the LK-MRS survey. 
For the targets in the {\sl K}2 fields, we chose footprints 
that are located as close as possible to the centers of the
{\sl K}2 campaigns but without overlap with the non-functioning CCD modules on the
{\sl Kepler} spacecraft. 
In contrast to the LRS plate classification into V/B/M/F plates {\citep[see details in the observations section of][and Z18b]{2015ApJS..220...19D},} there is only one {type of MRS}
plate for targets brighter than 15th magnitude in the Gaia $G$-filter. 
However, if the number of targets was not sufficient, we 
selected targets from the Gaia DR2 catalog to fill the remaining fibers. Therefore, the final input catalog for each footprint may also contain a few stars with magnitudes extending to $G\sim15.5$\, mag. 
Note that we decided to adopt Gaia $G$-band magnitudes (including $G_\mathrm{BP}$ and $G_\mathrm{RP}$) for all targets for homogeneity reasons. 
Figure\,\ref{fig1} shows the $G$ magnitude distribution of stars from the LK-MRS input catalog.

Table\,\ref{T1} lists the details of the 20 selected footprints. For each footprint, it contains the following columns: 
\\
(1) Plan ID: a string of 18 characters composed of the prefix ``TD'' (time domain), the middle part ``hhmmssNddmmss'' (the right ascension and declination of the central star truncated into seconds), and the postfix ``K01'' (the LK-MRS project); 
\\ 
(2) R.A. (2000): the right ascension of the central star at epoch J2000; 
\\ 
(3) Dec. (2000): The declination of the central star at epoch J2000; 
\\ 
(4) Target: the number of input targets; 
\\
(5) FS: the number of flux standard stars; 
\\ 
(6) Total: the total number of objects; 
\\
(7) KO: the number of objects cross-matched with the KIC/EPIC catalog for which {\sl Kepler/K}2 photometry is available; 
\\
(8) KNO: the number of objects cross-matched with the KIC/EPIC catalog for which no {\sl Kepler/K}2 photometry is available; 
\\
(9) NK: the number of objects not found in the KIC/EPIC catalog; 
\\ 
(10) Plate name: a string of 4 characters starting with a reference to the space mission (``K1''/``K2'') followed by a letter referring to the group (``a''/``b''/``c''/``d''/``e'') and an identification number (``1''/``2''/``3''/``4''); 
\\
(11) Field: reference to the location of the plates in the {\sl Kepler} field (``{\sl Kepler}'') or {\sl K}2 campaigns (``CNN'' with NN the campaign number).
\\
We note that the number of input targets in each plate is typically $\sim2000$ for a sparse target field (one single {\sl K}2 campaign) or $\sim3000$ for a dense target field ({\sl Kepler} or {\sl K}2 overlapping campaigns), whereas the number of the flux standard stars\footnote{Flux standard stars are used for flux calibration of LAMOST spectra. They are often selected by known stars with type of A or F.} is $\sim80$ for each plate. In total, we selected more than $54,000$ objects to be observed in a time span of 5 years starting in September of 2018. Almost all of the fibers are assigned to objects that are cross-matched to stars in the KIC/EPIC catalog, in particular to those for which high-quality space-based photometry is available ($\sim53\%$).

\startlongtable
\begin{deluxetable*}{cccrrrrrrcc}
\centering
\tablecaption{
Overview of the LK-MRS footprints after cross-match to the KIC/EPIC catalog. See details in text.
\label{T1}
}
\tablehead{
\colhead{Plan ID} & \colhead{R.A. (2000)}& \colhead{Dec. (2000)}& \colhead{Target}& \colhead{FS}&\colhead{Total$^b$} & \colhead{KO}& \colhead{KNO} &  \colhead{NK}  & \colhead{Plate name}       
 & \colhead{Field} }
\startdata
TD005004N074006K01   &  00:50:04.30   & +07:40:06.42   &  2065   &    78   &  2143   &  1146   &   992   &     5   &  K2b1   & C8   \\
TD005501N004722K01   &  00:55:01.40   & +00:47:22.40   &  2091   &    78   &  2169   &   970   &  1192   &    7   &  K2b2   & C8   \\
TD010142N094445K01   &  01:01:42.89   & +09:44:45.71   &  2212   &    79   &  2291   &  1157   &  1132   &     2   &  K2b4   & C8   \\
TD010605N031628K01   &  01:06:05.78   & +03:16:28.82   &  2136   &    78   &  2214   &  1069   &  1139   &    {6}   &  K2b3   & C8   \\
TD033722N181216K01   &  03:37:22.83   & +18:12:16.92   &  2729   &    77   &  2806   &   882   &  1918   &     6   &  K2c1   & C4   \\
TD035321N230725K01   &  03:53:21.17   & +23:07:25.00   &  2986   &    79   &  3065   &  1155   &  1907   &     3   &  K2c2   & C4   \\
TD043446N210613K01   &  04:34:46.90   & +21:06:13.43   &  2927   &    79   &  3006   &  1059   &  1940   &    7   &  K2c3   & C13   \\
TD045334N231856K01   &  04:53:34.10   & +23:18:56.41   &  3036   &    77   &  3113   &  1658   &  1450   &     5   &  K2c4   & C13   \\
TD082325N180811K01   &  08:23:25.83   & +18:08:11.20   &  2989   &    78   &  3067   &  1737   &  1325   &     5   &  K2d1   & C5C16   \\
TD084806N172341K01   &  08:48:06.36   & +17:23:41.19   &  2839   &    79   &  2918   &  2004   &   910   &     4   &  K2d3   & C5C16   \\
TD084844N123545K01   &  08:48:44.89   & +12:35:45.59   &  2845   &    78   &  2923   &  1658   &  1261   &     4   &  K2d2   & C5C16   \\
TD085754N225914K01   &  08:57:54.69   & +22:59:14.87   &  2670   &    80   &  2750   &  1418   &  1327   &     5   &  K2d4   & C5C16   \\
TD103356N023723K01   &  10:33:56.00   & +02:37:23.00   &  2300   &    80   &  2380   &  1006   &  1365   &     9   &  K2e2   & C14   \\
TD103827N055449K01   &  10:38:27.61   & +05:54:49.06   &  2186   &    75   &  2261   &   914   &  1343   &     4   &  K2e1   & C14   \\
TD104037N120443K01   &  10:40:37.32   & +12:04:43.24   &  2074   &    80   &  2154   &   805   &  1343   &     6   &  K2e3   & C14   \\
TD104844N081314K01   &  10:48:44.05   & +08:13:14.62   &  2090   &    75   &  2165   &   902   &  1256   &     7   &  K2e4   & C14   \\
TD190808N440210K01   &  19:08:08.34   & +44:02:10.88   &  3102   &    80   &  3182   &  2209   &    973   &   0   &  K1a3   & Kepler   \\
TD192102N424113K01   &  19:21:02.82   & +42:41:13.06   &  3186   &    80   &  3266   &  2391   &   831   &   44   &  K1a1   & Kepler   \\
TD192314N471144K01   &  19:23:14.82   & +47:11:44.87   &  3129   &    80   &  3209   &  2380   &    829   &   0   &  K1a4   & Kepler   \\
TD193637N444141K01   &  19:36:37.98   & +44:41:41.76   &  3123   &    80   &  3203   &  2282   &    920   &   1   &  K1a2   & Kepler   \\
\hline
Sum$^a$                  &  ~          & ~  &  52715   &    1570   &  54285   &  28802   &    25353   &   130   &  ~   & ~   \\
Fraction (\%)           &  ~       & ~  &  97.11   &    2.89   &  100   &  53.06   &    46.70   &   0.24   &  ~   & ~   \\
\enddata
\tablecomments{The cross-match identification to {\sl Kepler/K}2 targets is restricted to 3.7 arcsecs, same to the self-identification in \citet{zong2018b}.}
\tablenotetext{a}{A few overlaping targets in two different footprints are counted twice in the sum number. }
\tablenotetext{b}{The other fibers are assigned to sky light.}

\end{deluxetable*}

\subsection{Observations}
%--------------------------------------------------------------------------------
\begin{figure*}
\centering
\includegraphics[width=18cm]{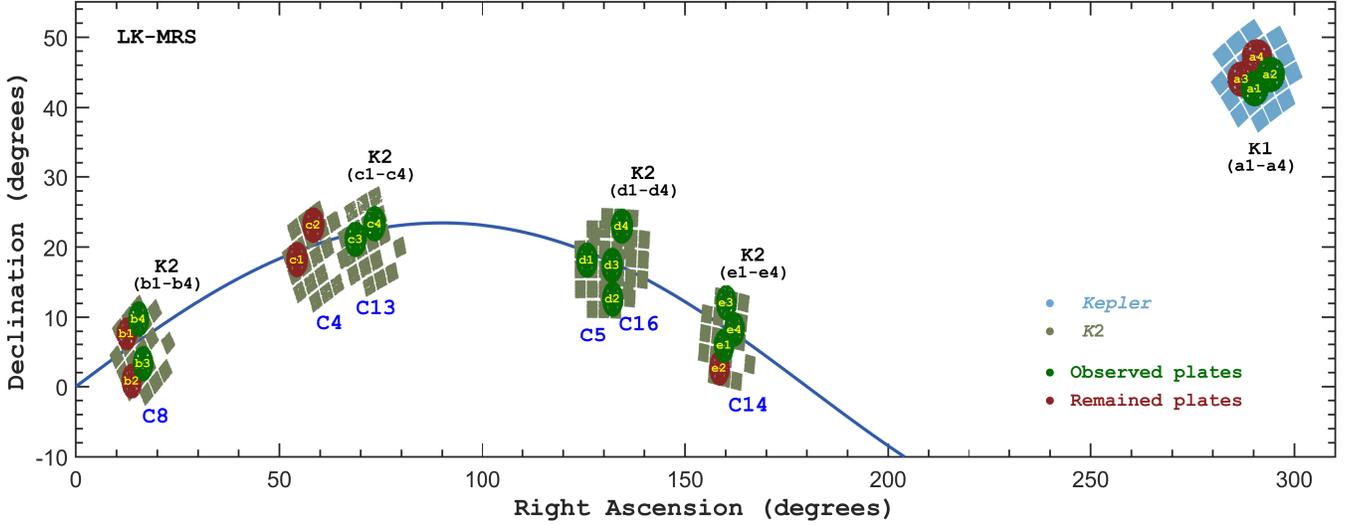}
\caption{Sky coverage of all footprints from the LK-MRS project stamped over the targets observed by {\sl Kepler} and {\sl K}2 campaigns. The solid line represents the ecliptic plane. The nomenclature of each plate is provided in the text. See Table\,\ref{T1} for detailed information and the exact location of each plate. 
\label{rd}}
\end{figure*}
%=================================================================================

\startlongtable
%%\begin{deluxetable*}{crcccc}
\begin{deluxetable*}{crccccrcc}
\centering
\tablecaption{Observation log of the LK-MRS project from 2018 September to 2019 June. \label{T2log}}
\tablehead{
\colhead{Plan ID} & \colhead{Exposure}& \colhead{Date}       & \colhead{Seeing} & \colhead{~~~~~}        & \colhead{Plan ID} & \colhead{Exposure} & \colhead{Date}       & \colhead{Seeing}             \\ 
\colhead{}        & \colhead{(s)}     & \colhead{yyyy/mm/dd} & \colhead{($^{\prime\prime}$)}& \colhead{}        & \colhead{}        & \colhead{(s)}      & \colhead{yyyy/mm/dd} & \colhead{($^{\prime\prime}$)}
}
\startdata
K2b4  & 1200 $\times$ 7  & 2018/10/17  & 3.3   & &  K2c3  & 1200 $\times$ 4  & 2019/01/19  & 3.4    \\ 
K2b3  & 1200 $\times$ 5  & 2018/10/19  & 2.9   & &  K2d3  & 1200 $\times$ 4  & 2019/01/20  & 4.5    \\ 
K2b3  & 1200 $\times$ 4  & 2018/10/24  & 2.4   & &  K2d3  & 1200 $\times$ 4  & 2019/01/23  & 4.0    \\ 
K2c3  & 1200 $\times$ 8  & 2018/10/24  & 2.5   & &  K2d3  & 1200 $\times$ 4  & 2019/01/24  & 4.0    \\ 
K2b3  & 1200 $\times$ 6  & 2018/10/28  & 6.8   & &  K2d3  & 1200 $\times$ 6  & 2019/01/25  & 3.3    \\ 
K2c4  & 1200 $\times$ 5  & 2018/10/30  & 2.8   & &  K2d1  & 1200 $\times$ 4  & 2019/02/11  & 3.5    \\ 
K2b3  & 1200 $\times$ 2  & 2018/11/16  & 4.0   & &  K2d4  & 1200 $\times$ 6  & 2019/02/13  & 3.8    \\ 
K2d2  & 1200 $\times$ 6  & 2018/11/21  & 2.8   & &  K2d4  & 1200 $\times$ 6  & 2019/02/21  & 2.7    \\ 
K2d4  & 1200 $\times$ 5  & 2018/11/25  & 3.7   & &  K2d4  & 1200 $\times$ 8  & 2019/02/23  & 2.8    \\ 
K2b3  & 1200 $\times$ 5  & 2018/11/26  & 3.0   & &  K2d4  & 1200 $\times$ {\bf 7}  & 2019/02/25  & 2.9    \\ 
K2b3  & 1200 $\times$ 4  & 2018/11/28  & 2.8   & &  K2d4  & 1200 $\times$ 5  & 2019/03/15  & 4.4    \\ 
K2d1  & 1200 $\times$ 8  & 2018/11/28  & 4.0   & &  K2e1  & 1200 $\times$ 6  & 2019/03/16  & 4.2    \\ 
K2d3  &  600 $\times$ 1  & 2018/11/29  & 3.7   & &  K2d1  & 1200 $\times$ 5  & 2019/03/18  & 3.0    \\ 
K2b3  & 1200 $\times$ 7  & 2018/11/30  & 2.8   & &  K2e1  & 1200 $\times$ 4  & 2019/03/21  & 3.7    \\ 
K2b3  & 1200 $\times$ 4  & 2018/12/13  & 3.1   & &  K2e1  & 1200 $\times$ 4  & 2019/03/23  & 3.6    \\ 
K2d3  & 1200 $\times$ 5  & 2018/12/17  & 2.8   & &  K2d4  & 1200 $\times$ 7  & 2019/03/24  & 2.6    \\ 
K2d1  & 1200 $\times$ 8  & 2018/12/19  & 2.7   & &  K2e4  & 1200 $\times$ 4  & 2019/03/24  & 2.6    \\ 
K2e3  & 1200 $\times$ 5  & 2018/12/19  & 2.6   & &  K2e1  & 1200 $\times$ 3  & 2019/04/25  & 5.2    \\ 
K2b3  & 1200 $\times$ 7  & 2018/12/25  & 2.8   & &  K1a1  & 1200 $\times$ 3  & 2019/05/21  & 3.1    \\ 
K2d1  & 1200 $\times$ 4  & 2018/12/27  & 6.0+  & &  K1a1  & 1200 $\times$ 4  & 2019/06/09  & 3.6    \\ 
K2e1  & 1200 $\times$ 7  & 2019/01/13  & 2.9   & &  K1a1  & 1200 $\times$ 3  & 2019/06/11  & 3.3    \\ 
K2c4  & 1200 $\times$ 4  & 2019/01/16  & 3.4   & &  K1a2  & 1200 $\times$ 4  & 2019/06/14  & 3.3    \\ 
\enddata
\tablecomments{
Four additional plates were observed as the testing program in three nights, whose central positions are same to those of K2d1, K2d2, K2d4 and K2e1 but with different fiber assignment. 
%Those plates are not included in the LK-MRS project but may have some targets overlaped.
Those plates are not included in the LK-MRS project but may have some targets in common. 
%{\bf Besides K1a1 had been observed 30 times during the testing phase in May 2018, see observational log in L19.}
Moreover, K1a1 has been observed 30 times during the testing phase in May 2018 (see Table\,1 of L19).
}
% \tablenotetext{a}{A few overlap targets in two different footprints are counted twice in the sum number. }

\end{deluxetable*}

During the transition period between the first and second phase of the regular survey (2017 September to 2018 June), tests were carried out with LAMOST equipped with medium-resolution spectrographs ($R\sim7500$). During that time, a plate named ``HIP95119'', located in the {\sl Kepler} field, was observed 30 times on five individual nights with exposure times ranging from 600\,s to 1200\,s \citep{2019RAA....19...75L}. 
Those exposures were the pioneering observations of the TD plates, as proposed later for the LK-MRS survey. That plate was also adopted for the LK-MRS project and was renamed with plan ID ``TD192102N424113K01'' or plate name ``K1a1''.

{LAMOST performs observations on both MRS and LRS programs since 2018 September. {The bright nights of each lunar month (from the 7th to the 22nd night) are scheduled for MRS observations, which are devoted to bright targets and are less affected by the sky brightness. The other nights are used for the LRS observations. }
}
According to the initial time allocation, about 3/8 and 5/8 of the MRS time is reserved to observe NT and TD plates, respectively. The exposure time is set to 20 minutes for (almost) all MRS plates. 
The NT plates will generally never be visited again after three exposures while the TD plates will each be visited about 60 exposures. 
{A Python code will first randomly decide the observing mode, NT or TD.
When TD mode is chosen}, the code randomly selects which kinds of plate, associated with one {(e.g., LK-MRS)} of the four parallel projects, will be the next one to be observed {\citep[see details in][]{2020arXiv200507210L}}. It takes the MRS time allocated to each of the projects into account. The initial probability for the LK-MRS project to be chosen is set to be $30\%$ among the four projects. 
Once a footprint has been observed, it will get a higher probability to be selected for future observations in order to collect 60 exposures as soon as possible, 
{or in another words, to finish the observation of that plate. As LAMOST can only observe the plates 2~hrs before and 2~hrs after their meridian passage, } 
the observations of a TD footprint will continue until the field leaves the LAMOST view in order to get more exposures for each pointing, which will save the overhead time. 
In practice, TD plates typically (but not strictly) begin with an observation time longer than 2~hrs, or 3+ exposures {(1 exposure = 20 min)} {plus the overhead time ($\sim 30$\,min) and read-out time (3 min)}. This action typically leads to a maximum of eight exposures for one pointing {(or observation)}. A detailed description of the procedure for the optimized selection of plates to be observed is given in \citet{2020arXiv200507210L}

\begin{figure}[!htp]
\centering
\includegraphics[width=8.5cm]{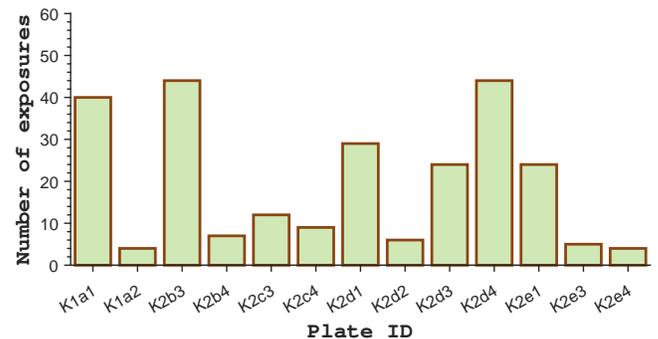}
\caption{
The distribution of the number of exposures for the 13 plates that have been observed in the LK-MRS survey so far.
\label{expo}
}
\end{figure}

With the above observation strategy, the LK-MRS footprints have been observed 223 times in 40 individual nights during the period from 2018 September until 2019 June, as summarized in Table\,\ref{T2log}. That corresponds to $\sim107$\,hrs of LAMOST time ($\sim74$\,hrs of exposure time, $\sim22$\,hrs of overhead time\footnote{In this estimation, an overhead time of 30\,min is taken. During the observations, it may be longer than 30\,min for a few plates.}, and $\sim11$\,hrs of read-out time). A total of 13 footprints have been visited in that period. Figure\,\ref{rd} shows their positions stamped on the {\sl Kepler/K}2 campaigns, along with the plates remaining to be observed. At least one plate has been observed in each of the five groups. Each group has a different optimal observing season. We clearly see that the observed plates are clustered around campaigns C5, C14, and C16, which is a consequence of the observation conditions being better in winter (longer nights, less clouds, calmer winds). In summer, during the monsoon season at Xinlong Observatory, LAMOST is undergoing maintenance. 
Figure\,\ref{expo} indicates how many times each of the observed 13 plates has been visited so far. These numbers range between 4 and 46. 
 
We note that the plates K2b3 and K2d4 have been observed more than 40 times, which is close to the number of exposures allocated to each plate in the first year of the LK-MRS survey\footnote{The initial plan was to obtain 48 exposures of five plates only in the period from 2018 September to 2019 June. 
However, in practice, the plates to be observed are chosen by the Python strategy program. In total, 13 different plates were selected in order to take the weather conditions and time allocations of all the active TD projects in parallel into account.}. There are four other plates which were observed 20-30 times (K1a1, K2d1, K2d3, and K2e1). For six plates, less than 10 exposures have been collected (K1a2, K2b4, K2c3, K2c4, K2d2, K2e3, and K2e4). In addition, there are four external plates that have been observed during bright nights during the testing of the MRS spectrographs. Those plates cover exactly the same region on the sky as four plates of the LK-MRS survey but the fibers were assigned to different stars within these fields\footnote{Their Plan IDs are TD084844N123545K02, TD082325N180811K02, TD103827N055449K02 and TD085754N225914K02, and they cover the same fields as the plates with names K2d2, K2d1, K2e1 and K2d4, respectively.}.

%-------------------------------------------------------------
% =============================================================

\subsection{Quality of the spectra}
% -------------------------------------------------------------
\begin{figure*}[!htp]
\centering
%\epsscale{.80}
\includegraphics[width=8cm]{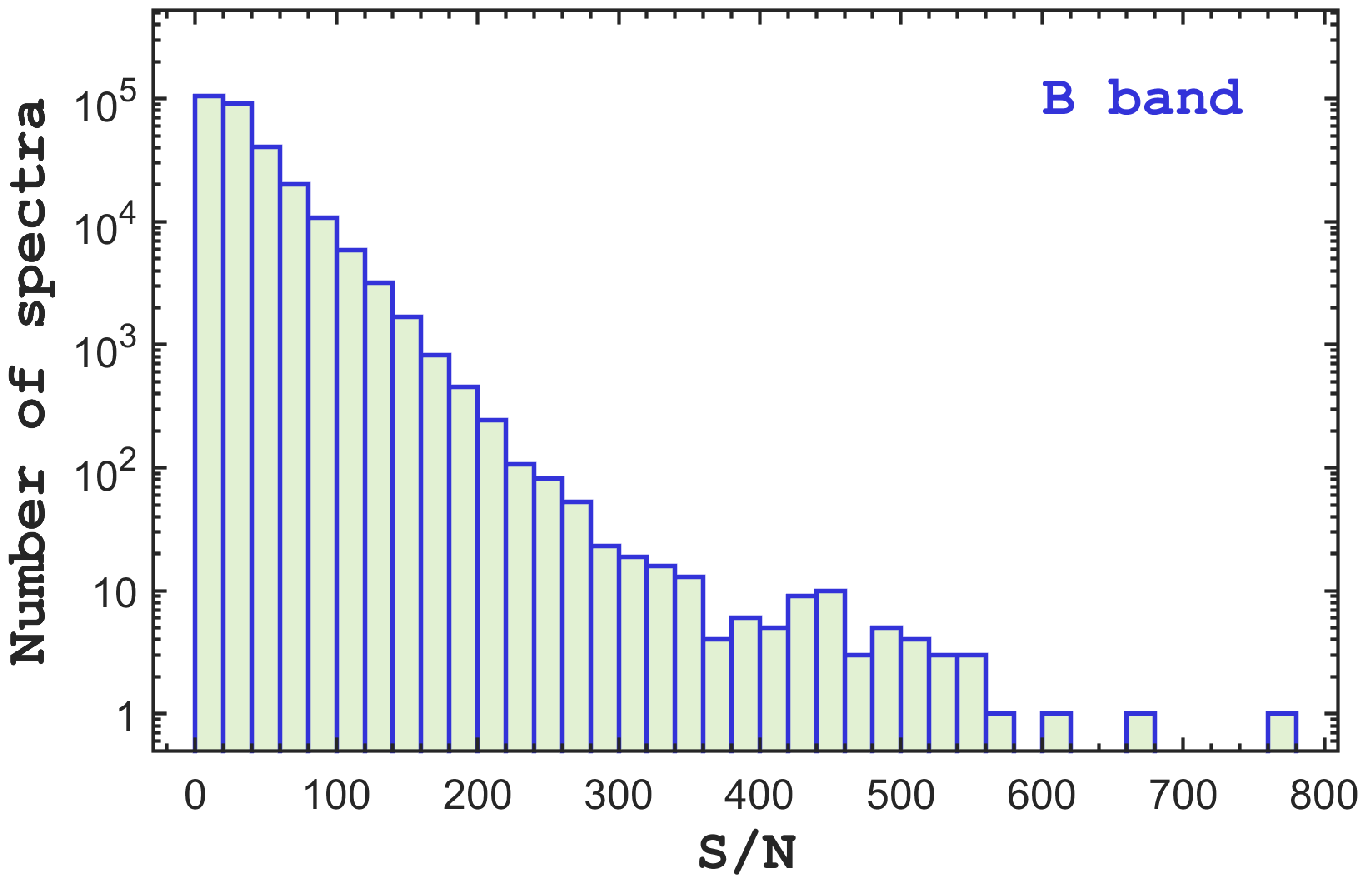}\hspace{0.5cm}
\includegraphics[width=8cm]{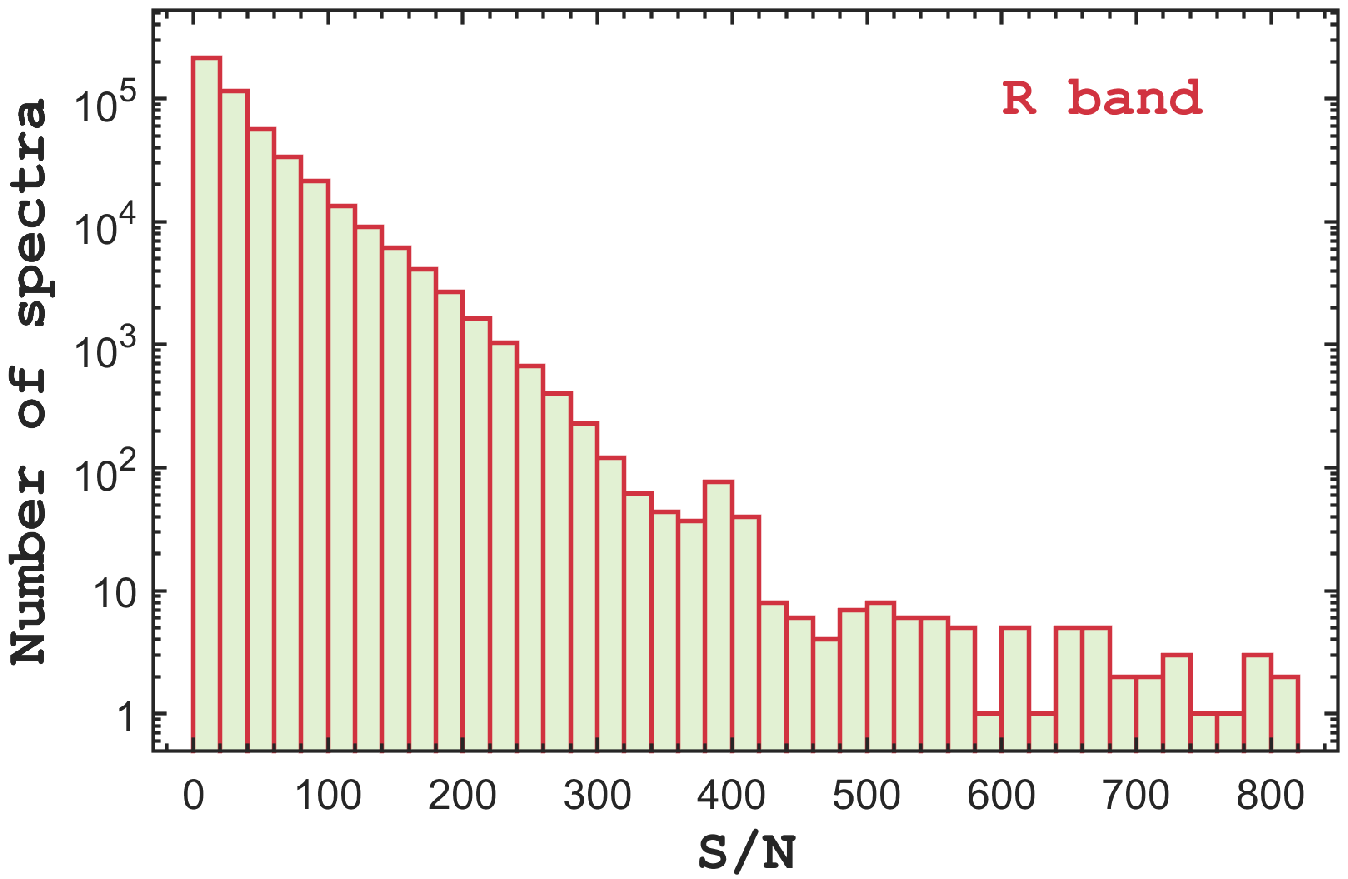}
\caption{
Distributions with a bin size of 20 in S/N in the blue (left) and the red (right) band for the high quality spectra obtained from the LK-MRS project. The vertical axis is in a logarithmic scale to make the small numbers at high S/N visible.
\label{figsnr}}
\end{figure*}
% ==================================================================

The first step in the reduction of the MRS data is the extraction of one dimensional (1D) spectra from two dimensional (2D) raw CCD frames. 
This process is similar to that of LRS spectra, except that the wavelength calibration is based on Th-Ar or Sc lamps without stacking of sub-exposures \citep[see details in][]{lamost2015,2019RAA....19...75L,2019ApJS..244...27W}. 
For each spectrum, the signal-to-noise ratio (S/N) per pixel was calculated at various wavelengths and the  median was taken as the final value. The LK-MRS survey collected 568,372 and 597,280 spectra in the blue and the red {arm} so far, respectively, including the spectra of the K1a1 plate observed in 2018 May and four test plates\footnote{
These plates are not observed in time allocated to the LK-MRS survey but they use the same input catalog. They have a different fiber assignment for part of the objects.}

However, due to some inoperative or inefficient fibers, a fraction of $\sim20\%$ of these spectra have a poor quality with S/N$<2$. 
We finally end up with 281,300 and 368,873 high-quality spectra (S/N$>$10) in the blue and red band, respectively. {The high-quality spectra in blue band  will be processed by the LASP pipeline \citep[see, e.g.,][]{2019ApJS..244...27W}}. 
Figure\,\ref{figsnr} shows the distribution of S/N for these spectra. We note that 175,661, 59,943, and 12,639 of the blue spectra have a S/N above 20, 50, and 100, respectively, corresponding to a fraction of $\sim62.5\%$, $\sim21.3\%$ and $\sim4.5\%$ of the high-quality MRS spectra. All the spectra collected in the LK-MRS survey will be made available to the public via LAMOST Data Release 7\footnote{http://dr7.lamost.org/} around 2021 September.

\begin{figure*}
\centering
%\epsscale{.80}
\includegraphics[width=18cm]{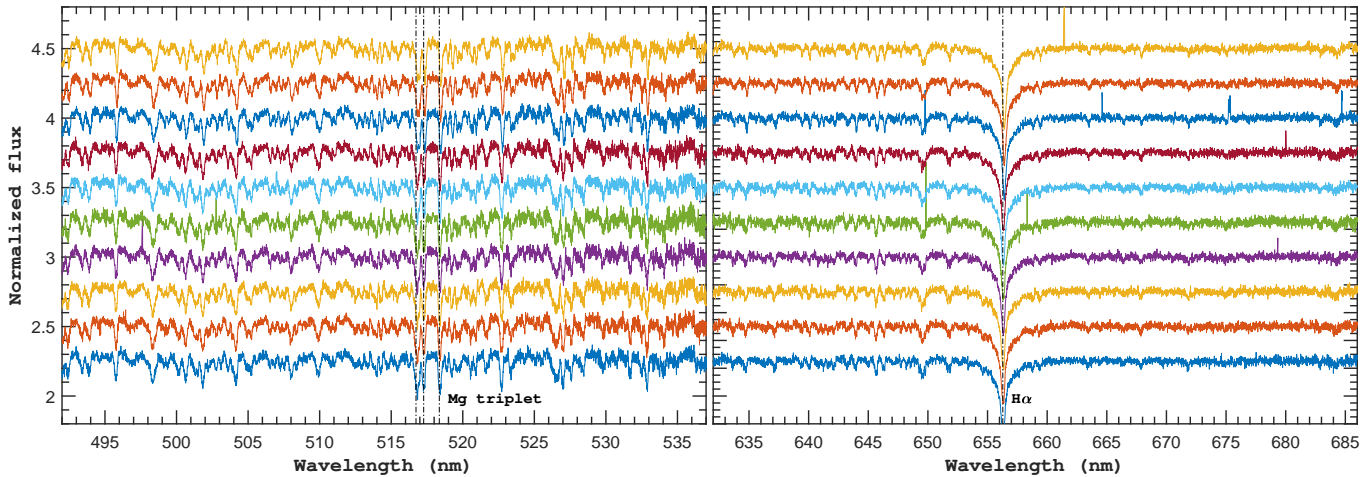}
\caption{
Examples of LAMOST medium-resolution spectra of KIC\,08685306, where the {segments in the blue arm (S/N$\sim50$) and in the red arm (S/N$\sim70$) are displayed} in the left and right panel, respectively. This time-series of ten spectra was obtained during two nights allocated to the LK-MRS survey. The flux is normalized and shifted for visibility reasons. The most distinct absorption lines (H$\alpha$ and the Mg\,{\sc i}b triplet) are marked with vertical lines.
}
\label{spectra}
\end{figure*}

Examples of high-quality MRS spectra of KIC\,08685306{, which is an eclipsing binary with a short orbital period (about 0.81 days) in the {\it Kepler} field \citep{2011AJ....141...83P},} are shown in Figure\,\ref{spectra}. These spectra were normalized using a third order polynomial fit discarding the outliers with $\sigma$ clipping: data points with residual fluxes above $+1\sigma$ or below $-3\sigma$ were removed, where $\sigma$ denotes the standard deviation of the residual flux. {They have S/N ratios of about 50 and 70 in the blue and red arms, respectively.} There are prominent absorption lines, like H$\alpha$ ($\lambda\sim656.3$\,nm) and the Mg triplet lines ($\lambda\sim517$\,nm), visible in the red and blue segments, respectively. We note that in the blue arm of the MRS spectra, the absorption lines of many other elements, including Fe\,{\sc i}, are clearly resolved. {A careful look at the time-series spectra shown in Fig.~\ref{spectra} allows one to detect shifts of the line centroids resulting from the orbital motion.} We note that only a selection of the observed spectra for KIC\,08685306 is presented, and that this star has been observed more than 30 times. For examples of raw spectra, we refer the reader to  \citet{2019ApJS..244...27W}.

%---------------------------------------------------------
\section{Properties of stellar parameters}
\subsection{Parameter catalog}
The stellar parameters are derived with a pipeline similar to LASP for LRS spectra, but adapted to the resolution $R\sim7500$ of the MRS spectra \citep{lamost2015,2019ApJS..244...27W}. 
Due to limitations of its template library, LASP provides the atmospheric parameters ($T_\mathrm{eff}$, $\log g$, and [Fe/H]) and radial velocities (RVs) only for stars with spectral types of late-A, F, G, and K. 
We note that the RV can be independently measured with other methods, even without obtaining the atmospheric parameters, such as the auto-correlation function method.  
Although the pipeline can provide the projected rotational velocity ($v\sin i$), its value is known to have poor accuracy, especially for slow rotators, because of the resolution of $R\sim7500$. { The results of v$\sin i$ from LASP is still under test by F. Zuo (in prep.) who will decide the cutoff value for the reliable v$\sin i$.}
We therefore do not provide the v$\sin i$ values here. {They will be presented and discussed in a future work based on the ROTFIT analysis of these spectra (A. Frasca et al., in preparation)}.
The abundances of $\alpha$-process elements [$\alpha/$Fe] can also be measured, but the quality of that parameter is still being investigated.
In the present form, LASP is applied to the blue arm spectra (495 -- 535\,nm) with, typically S/N$>10$, because this segment contains many more photospheric lines and, as consequence, provides better atmospheric parameters and RVs. However, a  
combined analysis of both segments will certainly improve the results, especially for the abundance determination of some elements for which lines of neutral and ionized species with different excitation potentials are present in the red and blue arm. A total of 281,300 spectra for 28,006 different targets {meet the requirements for this analysis.}

The LASP pipeline was successful for 258,979 entities, including a small number of spectra ($\sim3,000$) with $8<S/N<10$, resulting in atmospheric parameters and RV values for 21,053 targets\footnote{The current LASP version here contains a small fraction of spectra in this S/N range. They will be re-evaluated once the LASP code is updated for DR7. Based on the publication policy used in previous data releases, we expect that only results derived from LAMOST spectra with $S/N>10$ will be published.}. As most of these targets were visited at multiple epochs, we adopt the weighted average values for the stellar parameters of each target as: 
\begin{equation} \label{eqweight}
\overline{P} = \frac{\sum_k w_k \cdot P_{k}}{\sum_k w_k},
\end{equation}
where the index $k \in [1,N]$ is the sequence number of the measurement of parameter $P$ for one individual star. The weights $w_k$ are taken as the square of the S/N of the analysed spectrum. {\bf This arbitrary weighting criterion gives higher weight on the spectra with good quality but without omiiting the minor contribution from other spectra}.

However, the weighted RVs are corrected through a set of 2D systematic offset vectors, depending on the spectrograph and the epoch of the observation. This method was first introduced by L19. More details of this correction can be found in Appendix\,\ref{AppRV}. The origin of these complicated RV zero-point offsets is unclear {\bf but very possibly caused by the instrumental effects. During the observation, any (insignificant) changes affecting the optical systems may lead to the wavelength calibrating systems (as revealed by the precise position of spectra spanning on the CCD modules) varying even in a very tiny scale.} A similar phenomenon exists among high-precision RV measurements with a long time baseline \citep[see, e.g.,][]{2019MNRAS.484L...8T}. In Appendix\,\ref{AppAP}, we show that there is also a negligible offset effect occurring in the determination of the atmospheric parameters $T_\mathrm{eff}$, $\log g$, and [Fe/H].  

% ------------------------------------------------------------------
\begin{figure}
\centering
%\epsscale{.80}
\includegraphics[width=8cm]{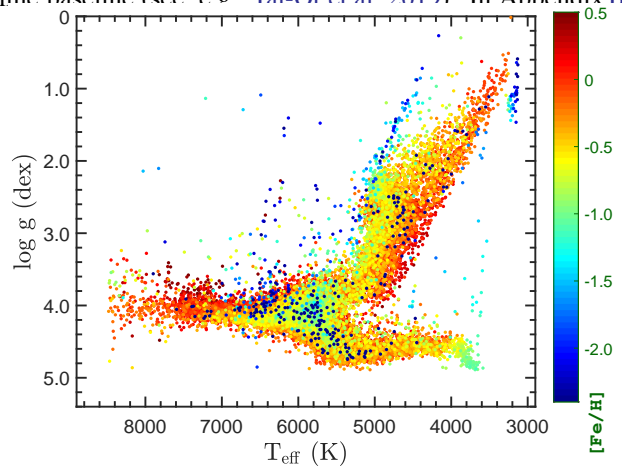}
\caption{Kiel diagram ($\log g$ versus $T_\mathrm{eff}$) of the 21,053 stars analyzed in the present paper.
The parameters are calculated as weighted average values from multiple measurements derived with the updated LASP pipeline. The points are color coded according to their value of [Fe/H].
The plotting sequence of the points was from high to low [Fe/H] values.
}
\label{swan}
\end{figure}
% ====================================================================

\startlongtable
\begin{deluxetable*}{cccrrrrrrc}
\centering
\tablecaption{
%Database of parameters and radial velocities from the first result of the MRSLKs survey. 
Database of atmospheric parameters and radial velocities obtained from the spectra collected in the first year of the LK-MRS survey. 
\label{T5}
}
\tablehead{
\colhead{Target name} & \colhead{KIC/EPIC} & \colhead{R.A. (2000)}& \colhead{Dec. (2000)}& \colhead{$T_\mathrm{eff}$}&\colhead{$\log g$} & 
\colhead{[Fe/H]} & \colhead{RV}& \colhead{Freq.} &  \colhead{Comment}   \\
\colhead{ } & \colhead{ } & \colhead{ }& \colhead{ }& \colhead{(K)}&\colhead{(dex)} & 
\colhead{(dex)} & \colhead{(km/s)}& \colhead{ } &  \colhead{ } 
  }
\startdata
... \\
J085535.16+223701.1 & 212129946 &  133.896528 &  22.616975 & 5900$\pm$24  &      4.09$\pm$0.04   &    0.02$\pm$0.02  &  -47.01$\pm$0.39   & 32 & \\
J085535.35+224553.5 & 212137266 &  133.897299 &  22.764873 & 4678$\pm$ 4  &      4.69$\pm$0.01   &   -0.06$\pm$0.01  &  -16.43$\pm$0.10   & 37 & \\
J085536.38+242400.3 & 212203673 &  133.901613 &  24.400088 & 5724$\pm$17  &      4.51$\pm$0.03   &   -0.35$\pm$0.02  &  -59.57$\pm$0.28   & 32 & \\
J085536.39+141257.6 & 211576681 &  133.901653 &  14.216021 & 5706$\pm$19  &      4.32$\pm$0.02   &   -0.13$\pm$0.02  &   20.82$\pm$0.21   & 5  & \\
J085536.41+241351.2 & 212198085 &  133.901736 &  24.230905 & 6135$\pm$77  &      4.15$\pm$0.09   &   -0.54$\pm$0.06  &   -3.44$\pm$0.16   & 8  & \\
J085536.51+122710.7 & 211453492 &  133.902158 &  12.452976 & 6540$\pm$47  &      4.14$\pm$0.04   &   -0.22$\pm$0.04  &   -3.56$\pm$0.18   & 5  & \\ 
J085536.53+153937.5 & 211681036 &  133.902242 &  15.660423 & 6038$\pm$110 &      4.03$\pm$0.11   &   -0.14$\pm$0.07  &   39.57$\pm$1.44   & 13 & \\
J085536.53+221948.5 & 212115652 &  133.902238 &  22.330144 & 4520$\pm$12  &      2.92$\pm$0.03   &   -0.03$\pm$0.02  &    4.56$\pm$0.25   & 37 & \\
J085536.58+133143.8 & 211527577 &  133.902457 &  13.528845 & 6048$\pm$67  &      4.31$\pm$0.10   &   -0.31$\pm$0.03  &  -13.10$\pm$0.32   & 5  & \\
J085536.62+135752.2 & 211558795 &  133.902588 &  13.964503 & 5073$\pm$16  &      2.96$\pm$0.05   &   -0.39$\pm$0.01  &   31.27$\pm$0.08   & 6  & \\
J085536.62+183750.9 & 211893502 &  133.902592 &  18.630812 & 4986$\pm$30  &      3.86$\pm$0.08   &    0.33$\pm$0.03  &   -2.02$\pm$0.58   & 7  & \\
J085536.93+223000.2 & 212124160 &  133.903910 &  22.500078 & 6291$\pm$120 &      4.12$\pm$0.10   &   -0.25$\pm$0.07  &   15.46$\pm$0.48   & 21 & \\
... \\
\hline
\enddata
\tablecomments{
%Only a few entities are shown and the entire catalog can be found through the online version. The cross-match identification to {\sl Kepler/K}2 targets is restricted to three arcsecs, same to that in \citet{zong2018b}. The error will be not given (default as zero) if the Freq number is one. The detailed calculation of these parameters and RVs can be found in text.
Only a few entries are shown here. The entire catalog is available in the online version of this paper. The cross-match identification to {\sl Kepler/K}2 targets is restricted to 3.7 arcsecs (cf. \citealt{zong2018b}). In the case that only one spectrum is available a star (see the column "Freq."), the errors are set to the uncertainty derived by LASP. A detailed description of the calculation of the values of the atmospheric parameters, RVs, and their errors is given in the text.
}

\end{deluxetable*}

Table\,\ref{T5} contains the full catalog of the 21,053 analysed stars from the LK-MRS survey up to 2019 June.
It is composed of the following columns: 
\\
(1) Target name: the LAMOST input ID or the name of the LAMOST target; 
\\ 
(2) KIC/EPIC: the cross-match identification to the KIC/EPIC catalog where a coordinate separation of 3.7 arcsec\footnote{The nearest star is chosen if more than one star is identified. An enlargement or decrease of the maximum separation distance does not change the results of the cross-matching significantly.} is used as the limit (if available); 
\\
(3) R.A. (2000): the observed right ascension (epoch J2000.0) of the fiber in degrees; 
\\
(4) Dec. (2000): the observed declination (epoch J2000.0) of the fiber in degrees; 
\\
(5) $T_\mathrm{eff}$, (6) $\log g$, (7) [Fe/H], (8) RV: the weighted average parameters 
from Equation\,(\ref{eqweight}) and with their {standard errors calculated as:}
\begin{equation}
\sigma_w(\overline{P}) = \sqrt{\frac{N}{N-1}\frac{\sum_k w_k \cdot (P_{k} - \overline{P})^2}{\sum_k w_k}},
\end{equation}  \\
(9) Freq.: the number of MRS spectra that were analysed for this target; 
\\
(10) Comment: extra information about the star or the analysis (if necessary).

Figure\,\ref{swan} illustrates the location of the 21,053 analysed stars in a Kiel diagram ($T_\mathrm{eff}$ vs. $\log g$), with an extra dimension for [Fe/H] through a colormap. Similar to the results obtained from the LRS spectra in the LK-Project, $T_\mathrm{eff}$ is mainly found in the range [4000, 7000]~K while $\log g$ is found between 5 and 1\,dex. Most stars show close-to-Solar metallicities, as indicated by the red points. It is clear that most stars are located in either the main sequence or the red giant branch. We note that the giant branch displaces towards cooler temperatures as the metallicity increases, in line with the predictions of stellar evolution theory \citep[see, e.g.,][]{2000ApJ...543..955B,2015RAA....15..549Z}.

\begin{figure*}
\centering
%\epsscale{.80}
\includegraphics[width=8cm]{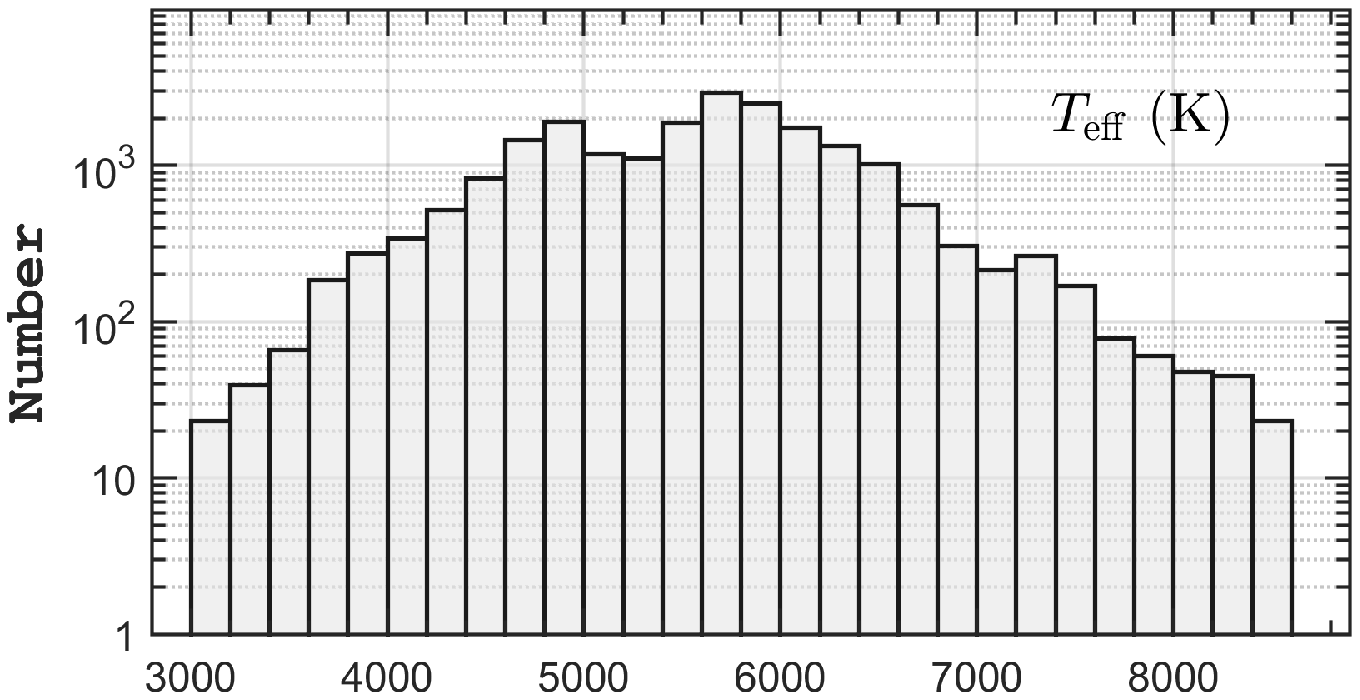}\hspace{0.5cm}\includegraphics[width=8cm]{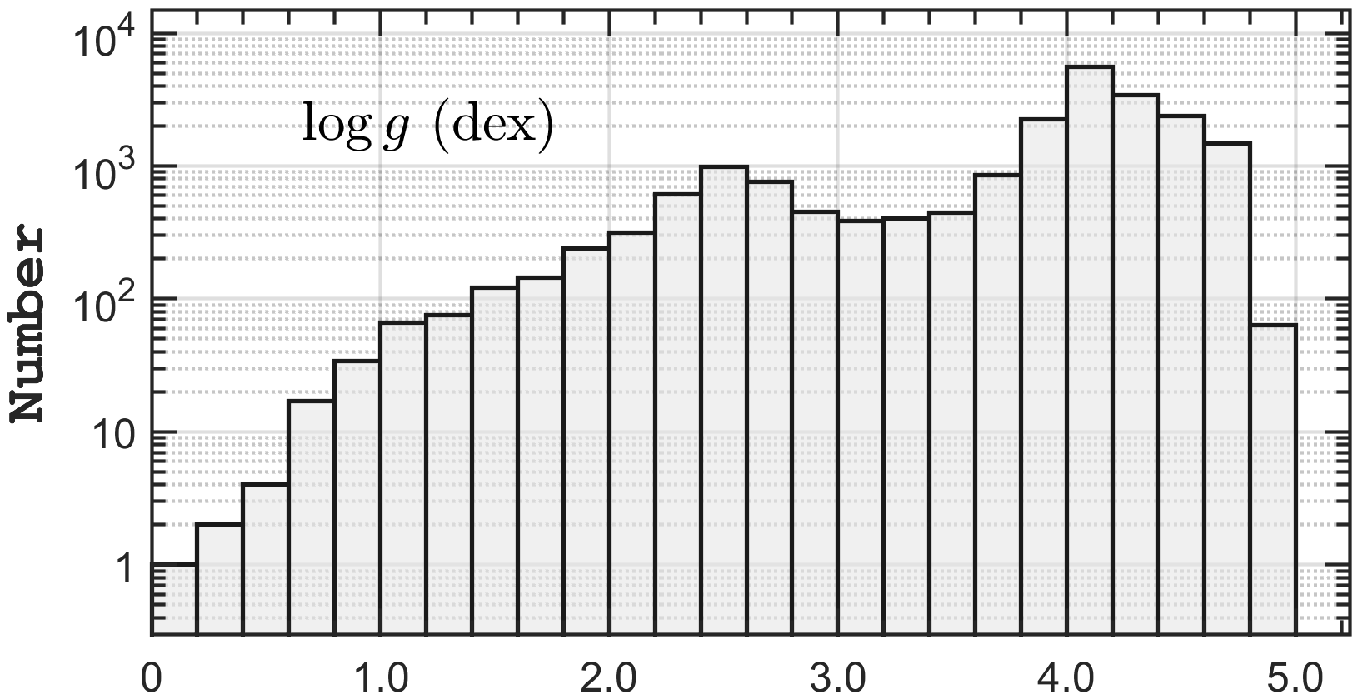} \\\includegraphics[width=8cm]{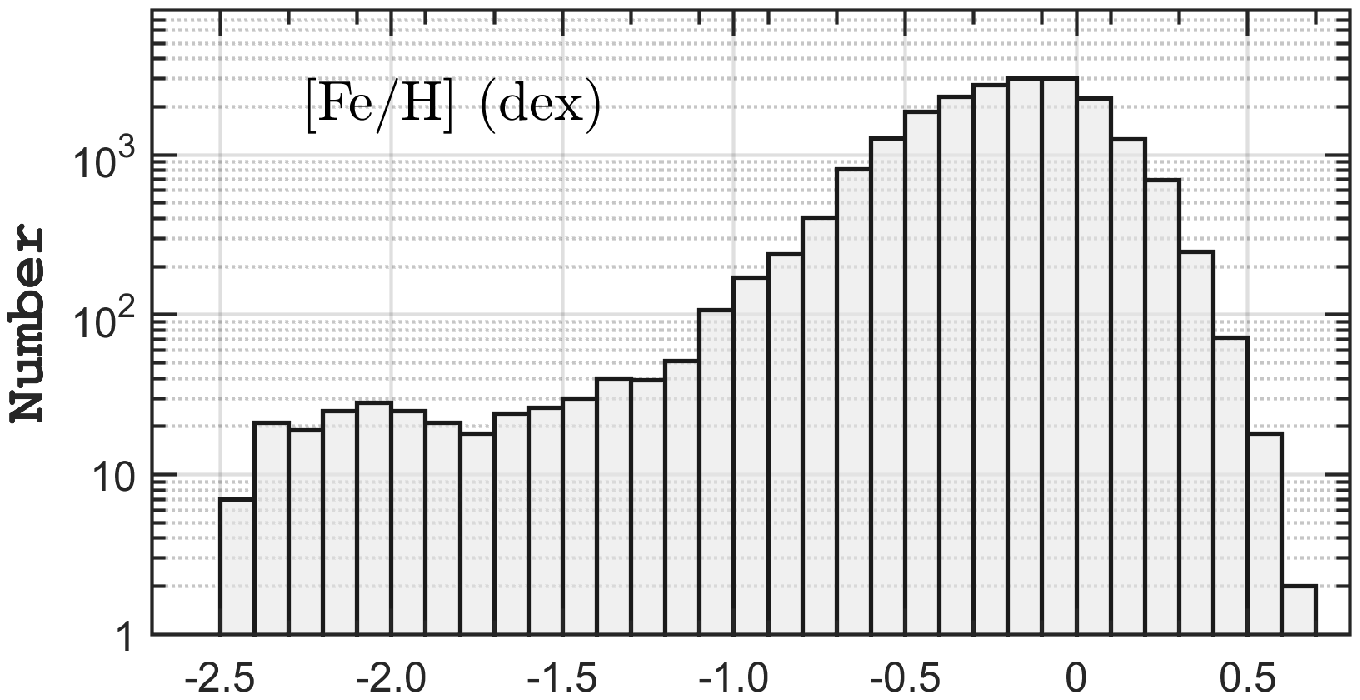}\hspace{0.5cm}\includegraphics[width=8cm]{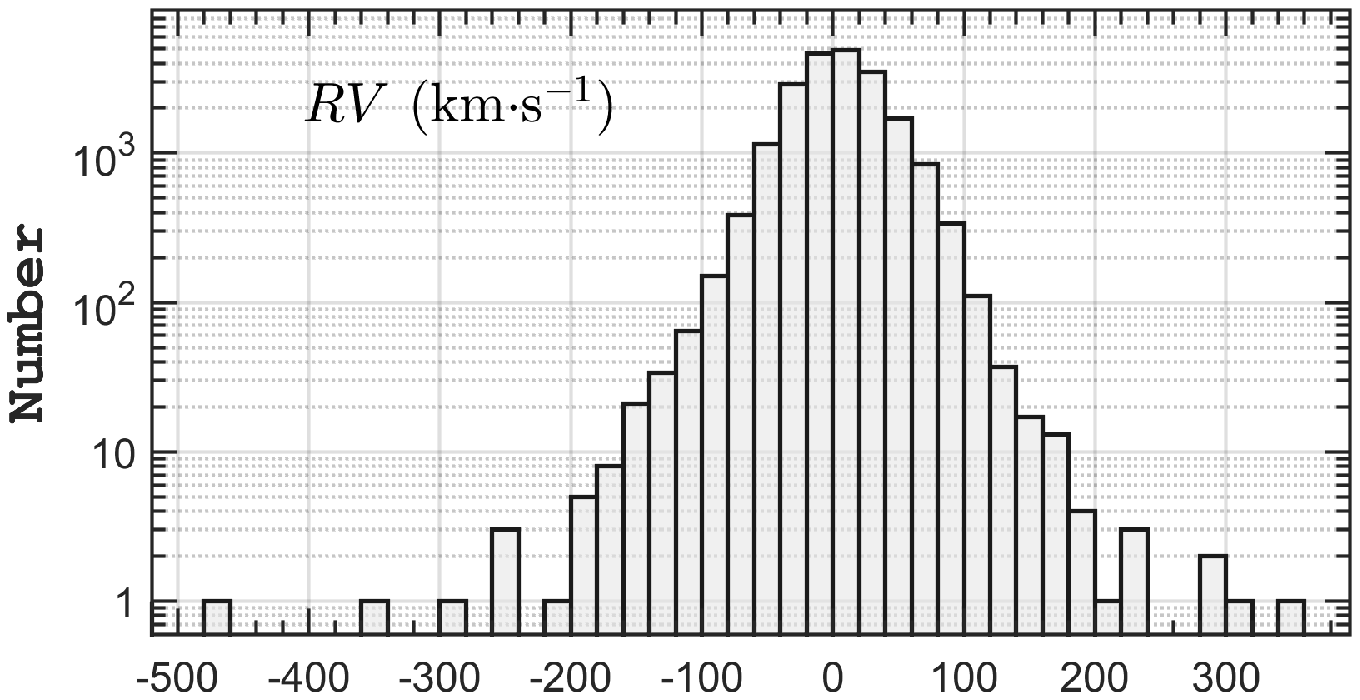}
\caption{Histograms of the weighted average values of the atmospheric parameters derived for the 21,053 targets. Top left: the effective temperature $T_\mathrm{eff}$ (K, bin size of 200\,K). Top right: the surface gravity $\log g$ (dex, bin size of 0.2\,dex). Bottom left: the metallicity [Fe/H] (dex, bin size of 0.1\,dex). Bottom right: the radial velocity RV (km\,s$^{-1}$, bin size of 20\,km\,s$^{-1}$).}
\label{fours}
\end{figure*}

Figure\,\ref{fours} displays the histograms of the weighted average values of $T_\mathrm{eff}$, $\log g$, [Fe/H], and {\sl RV} for the entire catalog. A bimodal distribution is visible in the $T_\mathrm{eff}$ histogram, with peak values near $\sim4800$ and $5800$\,K, caused by the projection of the giant and the main sequence stars in Figure\,\ref{swan}, respectively. 
The cut-off values of $T_\mathrm{eff}$ are 3200 and 8500\,K, corresponding to the current limits imposed by the LASP pipeline. 
{\bf However, it should be carefully when one uses temperature of the target near the two limits ($T_\mathrm{eff} > 7500$\, or $T_\mathrm{eff} < 3500$\,K) where LASP does not work so well as in the range of $T_\mathrm{eff} \in [3500, 7500]$\,K.}
A similar bimodal distribution also occurs for $\log g$, with peaks at $\sim$2.5 and $4.2$\,dex. 
Most of the analysed objects have [Fe/H] values spanning from $-0.9$ to 0.4\,dex, with the solar value occurring most frequently. 
Objects with [Fe/H] $<-1.5$ show a different distribution compared to that of the LK project, where a logarithmic decrease in number was found (Z18b). We note that the MRS spectra cover a relatively short wavelength range, have a higher resolution, and represent a much smaller data sample than that of the LK project. 
This could have played a role in the different [Fe/H] distributions. However, this difference may be real, if we consider the different magnitude limits of the two surveys, which explore different volumes of the solar neighbourhood.
In the distribution of the RV values, the highest peak occurs around $\sim0$\,km\,s$^{-1}$. 
There are a few stars with $|RV|>300$\,km\,s$^{-1}$. They are classified as candidate high-velocity stars. 
We note that the unimodal RV distribution shown in Figure\,\ref{fours} is somewhat different from that of Z18b; this may also be a 
consequence of the different sample sizes.

% ------------------------------------------------------------------
% ====================================================================

\subsection{Measurement uncertainties}
Unlike the LK-project, the LK-MRS survey collects spectra at different epochs, which gives us the ability to evaluate the internal uncertainties through the differences between multiple measurements of the same object. 
{This provides a unique opportunity to} assess the general performance of MRS spectroscopic observations of LAMOST. 
We used the method of the unbiased estimator, where the uncertainties are based on the differences calculated with the formula:
\begin{equation}
\Delta P_k = (P_k - \overline{P}) \cdot \sqrt{N/(N - 1)}.
\end{equation}
The uncertainty distribution of the parameters $T_\mathrm{eff}$, $\log g$, [Fe/H], and RV, along with their S/N, are shown in Figure\,\ref{err}. 
We clearly see that the precision of the measurements improves as their S/N increases. 
A small fraction of the points are outliers, which might be the measurements obtained for variable stars, in particular those variable in RV. 
{\bf Or concretely, the LASP pipeline automatically treats each spectrum as from one single star even in the case of binary star.
}
To evaluate the uncertainties correctly, we first discard outliers by applying $3\sigma$-clipping to the differences 
for S/N intervals with a bin size of 6. Typically, this process is iterated two or three times until the number of remaining points does not change significantly. 
Next, the distribution of the remaining differences for each parameter $P$ was fitted with a power law of the form:
% ------------------------------------------------------------------
\begin{figure*}
\centering
%\epsscale{.80}
\includegraphics[width=18cm]{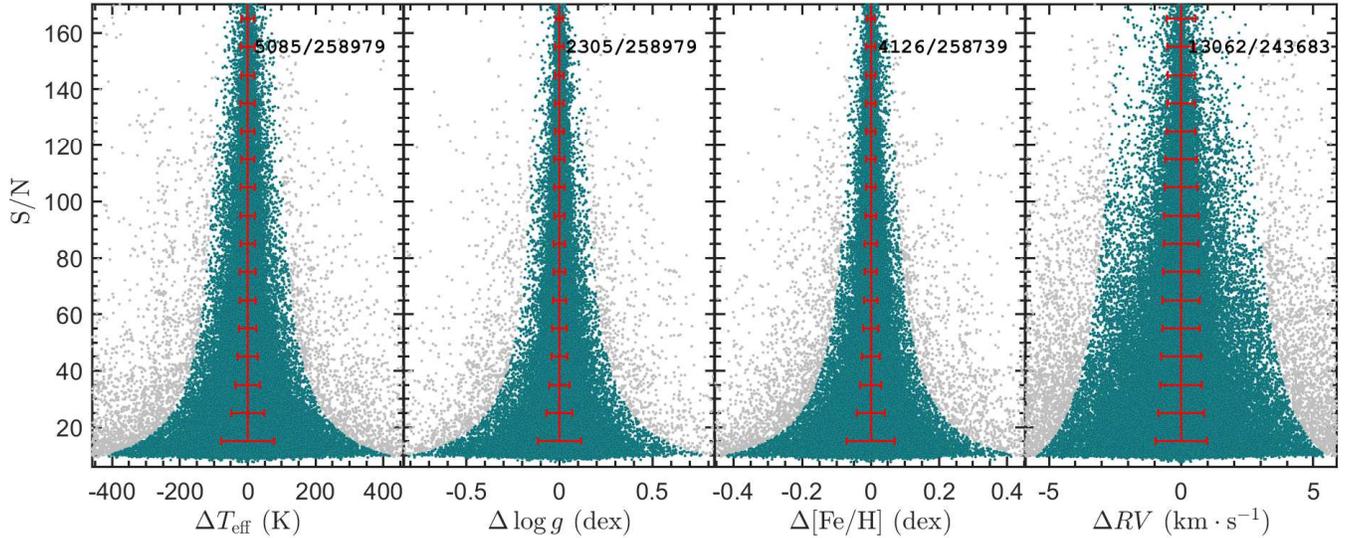}
\caption{Internal uncertainties of $T_\mathrm{eff}$, $\log g$, [Fe/H] and RV as a function of the quality of the MRS spectra (S/N). 
The grey and green dots refer to the outliers and remaining data points for the estimation of those uncertainties, respectively. The size of these samples is given on the top right corner in each panel.
See text for more details.
The error bars, stamped on the zero-point (vertical lines), are the estimated $1\sigma$ internal uncertainties for S/N$= 16, 26,...,166$. 
}
\label{err}
\end{figure*}
% ------------------------------------------------------------------
\begin{equation} \label{eqpower}
\sigma_\mathrm{P} = a\cdot x^{b} + c,
\end{equation}
where $x$ denotes the value of S/N $\in [0, 300]$. 
{The measurements with $S/N > 300$ are not used because there are too few of them.}
From this equation, the quantity $\sigma_\mathrm{P}$ defines the general uncertainty of the atmospheric parameters and RV for one spectrum as a function of its S/N, which is not same as $\sigma_w(\overline{P})$ (the uncertainty for one star through multiple measurements). 
Afterwards, we excluded the final outliers, which are those data points for which the $\Delta P$ value is larger than {$6\sigma_\mathrm{P}$} (grey in Figure\,\ref{err}). 
The choice of the factor 6 prevents too many data points from being considered as final outliers. The total number of outliers amounts to 
less than $2\%$ for all the parameters except for RV, for which it is about $5\%$.
This is larger than expected for a normal distribution of uncertainties, and it is likely caused by sources with a genuine RV variation (pulsating stars and binaries) in the sample. 
We note that for RV, the initial number of data points is about 15,000 lower compared to those for the atmospheric parameters. They are removed during the procedure to correct for zero-point offsets (see Appendix\,\ref{AppRV} or L19) 
{due to the the following reasons. To correct the systematic offsets, the plates need common constant (CC) stars, which means that 1) those stars must have spectra in every exposures of those plates and 2) their RV scatters must be less than a certain value (e.g., 1.0 km/s) that can be assumed as ``constant''. Therefore, a plate needs to contain enough spectra to serve as CC stars otherwise the calculated value of offset is less reliable. For instance, in the poor weather condition only a few stars selected as CC stars, the risk of these stars is not RV constant will be increase as their weight is very large. In practice,} each plate needs to contain at least 540 {spectra that can have enough} constant RV stars. {A plate containing spectra less than 540 were discarded during the RV correction procedure.} In addition, there are two plates for which the RV measurements from one particular spectrograph are inconsistent with each other with a very large scatter.\footnote{Those data can be easily identified through the method we provided in Appendix.} We do not consider those two sets of spectra in the estimation of the RV uncertainties.

The final values for the uncertainties $\sigma_{\rm P}$ are estimated, again, with equation\,(\ref{eqpower}), which is fitted to the sample of remaining differences for each parameter $P$. For the fits, 80 discrete data points, starting at S/N = 8 with a step of 2, were used. 
{Note that this fitting is not the same as the fitting function used for the clipping of outliers in their S/N range.}
In Figure\,\ref{err}, we show 16 of {of the final uncertainties}, starting at S/N = 16 with an interval of 10. Table\,\ref{abc} lists the values of the coefficients $a$, $b$, and $c$ for the optimal power law fit for each parameter. With these fitting coefficients, we calculated the uncertainties at S/N\,=\,10, 20 and 50. They are also listed in Table\,\ref{abc}. These values are similar to those of other studies. This is particularly true for the error estimation of RV, where L19 and \citet{2019ApJS..244...27W} found an internal uncertainty estimate of  $\sim1$\,km\,s$^{-1}$ for RV-values derived from MRS spectra with a S/N of the order $10 \sim20$.

Note that we may overestimate the uncertainty of RV for S/N$>40$ (or even lower). At these high S/N-values, there are several potential outliers that can not be directly traced with a simple power law fit. These data points have a very high probability of originating from either binaries or RV variables. If they were removed, the estimated precision of RV would approach $0.3\sim0.5$\,km\,s$^{-1}$ for ${\rm S/N}>50$, as documented in L19. A better estimate of the errors would be obtained if those RV variables could be removed before the analysis.

\begin{deluxetable}{lccr|ccc}
\centering
\tablecaption{
The values of the coefficients $a$, $b$, and $c$ of the optimal fit with Equation\,\ref{eqpower} to determine the internal uncertainties for the atmospheric parameters and RV. The results for S/N = 10, 20, and 50 are given in the columns on the right.
\label{abc}
}
\tablehead{
\colhead{ } & \multicolumn{3}{c}{Fitting coefficients} & \multicolumn{3}{c}{S/N} \\
\colhead{ } & \colhead{$a$}& \colhead{$b$}& \colhead{$c$}& \colhead{10} & \colhead{20} & \colhead{50}}
\startdata
$T_\mathrm{eff}$~~~~~~~~~~(K) & 1109   &   -1.105  & 14.24  & 101     & 55   & 29   \\
$\log g$          ~~~~(dex) & 1.241  &   -0.951  & 0.012  & 0.15    & 0.08 & 0.04 \\
$[\mathrm{Fe/H}]$ ~(dex) & 0.893  &   -1.032  & 0.008  & 0.091   & 0.048 & 0.024 \\
RV             ~~~~~~(km\,s$^{-1}$) & -0.420  &   -0.201  & 1.669  & 1.00   & 0.90 & 0.75 \\
\enddata
\tablecomments{
Contrary to the atmospheric parameters, the coefficient $b$ for RV is far from -1, a value which indicates that the precision is proportional to the inversion of S/N. Note that a reciprocal fit to RV was applied by L19. This difference is probably the result of the increase in the number of undefined outliers of RV when S/N increases.
}
\end{deluxetable}

\section{Comparison with other surveys}
The LK-MRS survey aims to build a database that hosts a very large sample of stars for which MRS spectra are collected at multiple epochs. 
It is the first large project that is dedicated to make possible the combination of time-series of space-based ultra-precise photometric data with time-series of ground-based spectra in order to perform in-depth studies in stellar physics.
{Although the good internal precision of the measurements of atmospheric parameters and RV has been assessed in Section~3, it is necessary to evaluate their accuracy by comparison to other large surveys, as an external quality control.} Moreover, the derived parameters are calculated from the {blue-arm spectra only, which have a relatively short wavelength coverage.}
Whether this may lead to large discrepancies or not needs to be checked. We therefore use the results obtained from the LAMOST LRS, APOGEE and GAIA surveys as external calibrators. These are the only large spectroscopic surveys that have enough targets in common with the LK-MRS survey to allow a statistically significant comparison of the results. For all comparisons, we use the weighted average values for each star in the LK-MRS survey, instead of the multiple individual measurements for the 21,053 stars listed in Table\,\ref{T5}.

\subsection{LAMOST low-resolution survey}
% ------------------------------------------------------------------
\begin{figure*}
\centering
%\epsscale{.80}
\includegraphics[width=6cm]{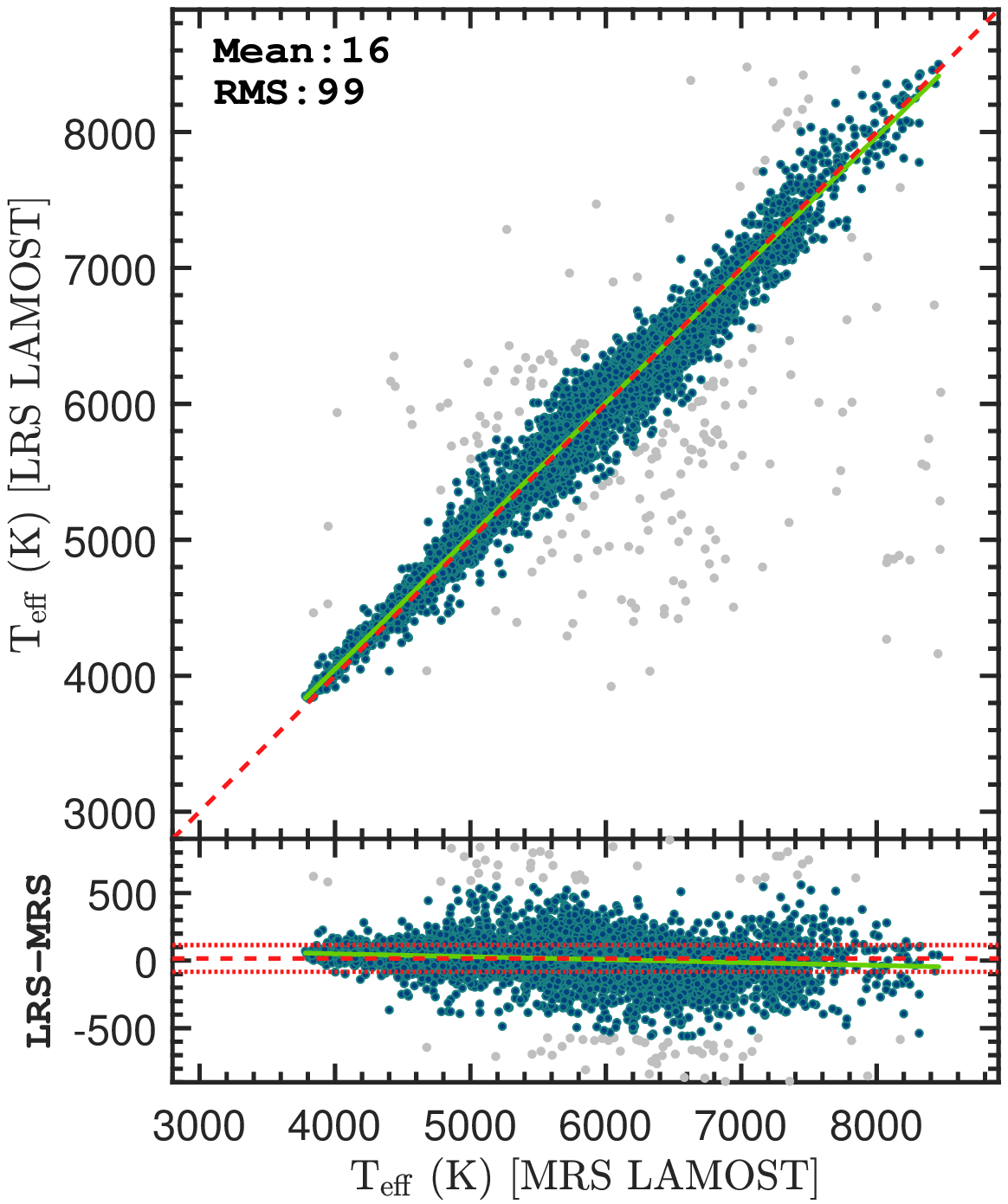}\includegraphics[width=6cm]{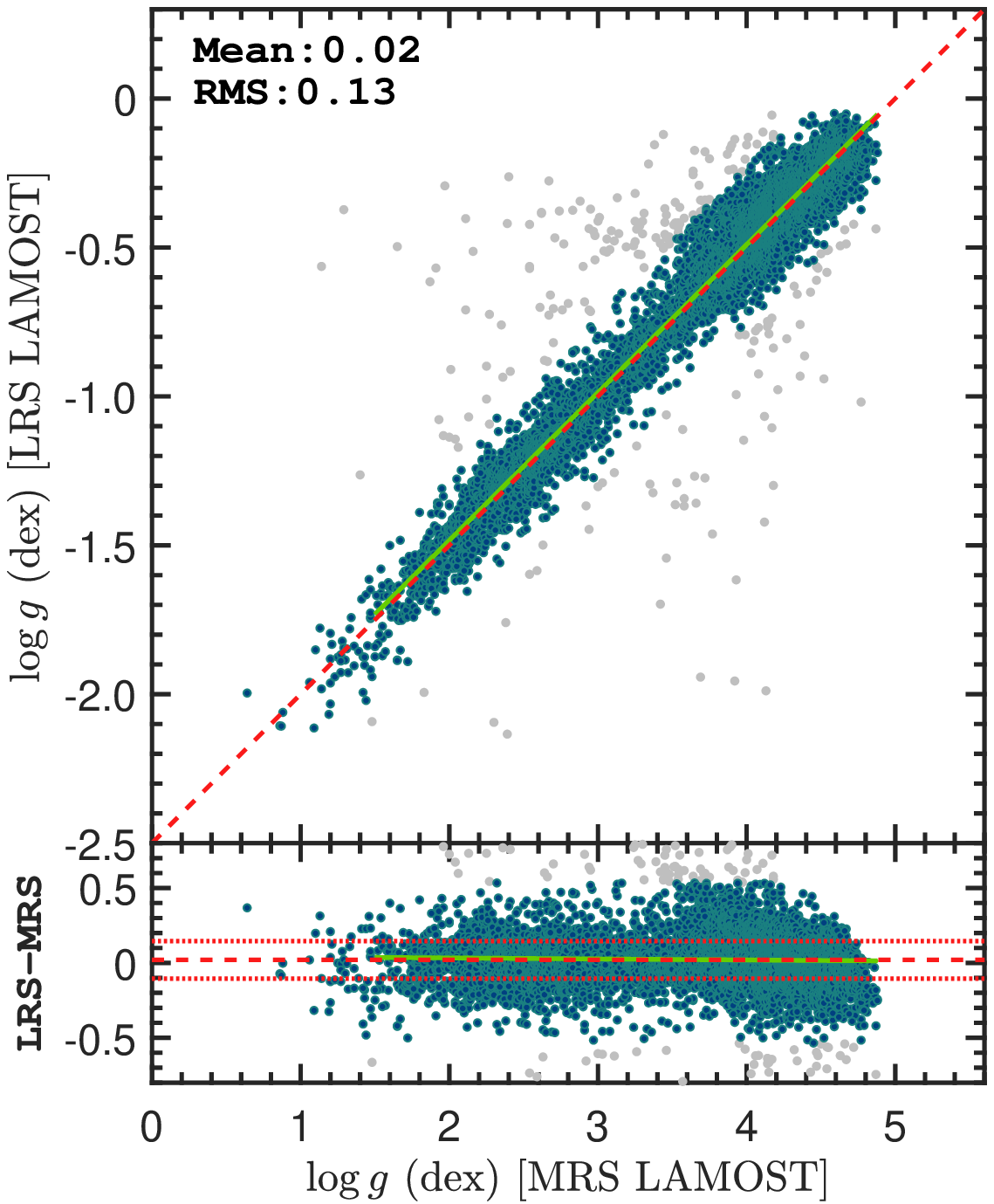}\includegraphics[width=6cm]{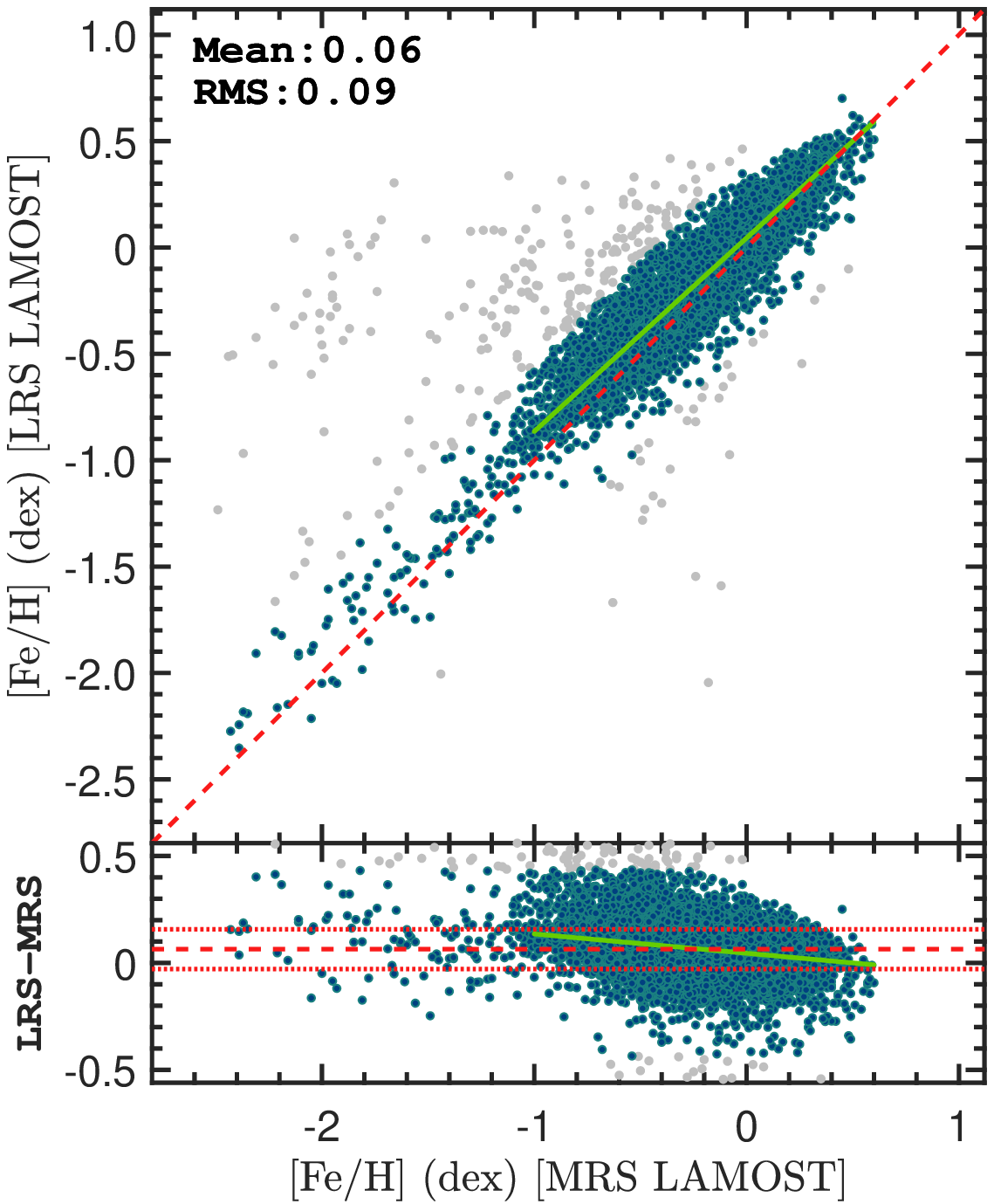}
\caption{Comparison of the $T_\mathrm{eff}$ (left), $\log g$ (middle) and [Fe/H] (right) values for the 14,997 stars in common between the LK-MRS and LAMOST LRS surveys. {The solid lines show the linear regression for the comparisons, limited to the parameter ranges spanned by the majority of the points.}
The dashed lines represent the bisectors (top) and the mean values in the residuals (bottom panels) with their associated $\pm$1$\sigma$ deviations (dotted lines). The pale grey points are the $3\sigma$ outliers from {the means}. 
}
\label{lmrs}
\end{figure*}
% ====================================================================

LAMOST DR7 contains more than 11 million high-quality LRS spectra. 
Apart from the gratings, the MRS instrument shares the other components with the LRS spectrographs. 
After a cross-match with the LRS of LAMOST DR7 by using a maximum distance separation criterion of 3 arcsecs, a total of 14,997 targets are found to be in common. 
This means that a fraction of $\sim70\%$ of the MRS targets were also observed by the LRS survey. 
Figure\,\ref{lmrs} shows the comparison of the atmospheric parameters between those two catalogs. In general, the values of $T_\mathrm{eff}$, $\log g$, and [Fe/H] are found to be consistent with each other. 
After removing the outliers with $3\sigma$ clipping, the standard deviations of the residuals (RMS) of the atmospheric parameters are very similar to the internal uncertainties found for MRS spectra with ${\rm S/N}=10$ (cf. Table\,\ref{abc}). There are small offsets between the two catalogs, as indicated by the mean values of the residuals (16\,K for $T_\mathrm{eff}$; 0.02\,dex for $\log g$; 0.06\,dex for [Fe/H]). 
The offset in the metallicity is comparable to its RMS, i.e. 0.06\,dex vs. 0.09\,dex. {The solid lines in Figure\,\ref{lmrs} represent linear
regression fits for all three parameters after the 3$\sigma$ outliers have been excluded. 
{\bf Those outliers are very possibly coming from the results of inappropriate measurement due to, for instance, binary stars, variable stars, or mistakes by parameter templates.
They typically account a few (1 or 2) percents of the entire sample.} In the cases of $T_\mathrm{eff}$ and $\log g$, the agreement with the bisector (dashed) lines are excellent.  In the case of [Fe/H] however, the linear regression line has a slope {of 0.92 which is} significantly smaller than unity indicating a systematic difference from the LRS results.}

% ====================================================================
\subsection{APOGEE}
\begin{figure*}
\centering
%\epsscale{.80}
\includegraphics[width=6cm]{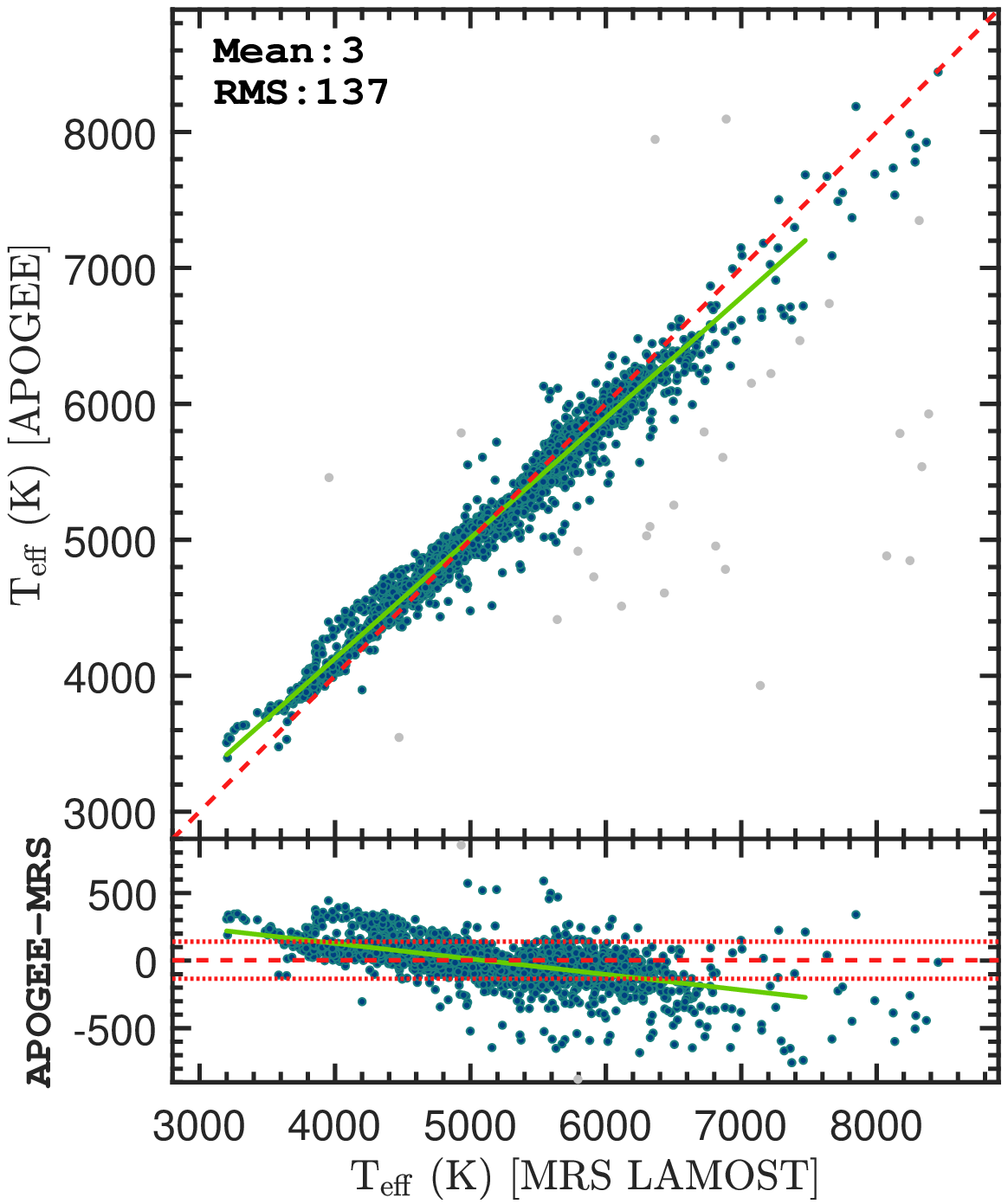}\includegraphics[width=6cm]{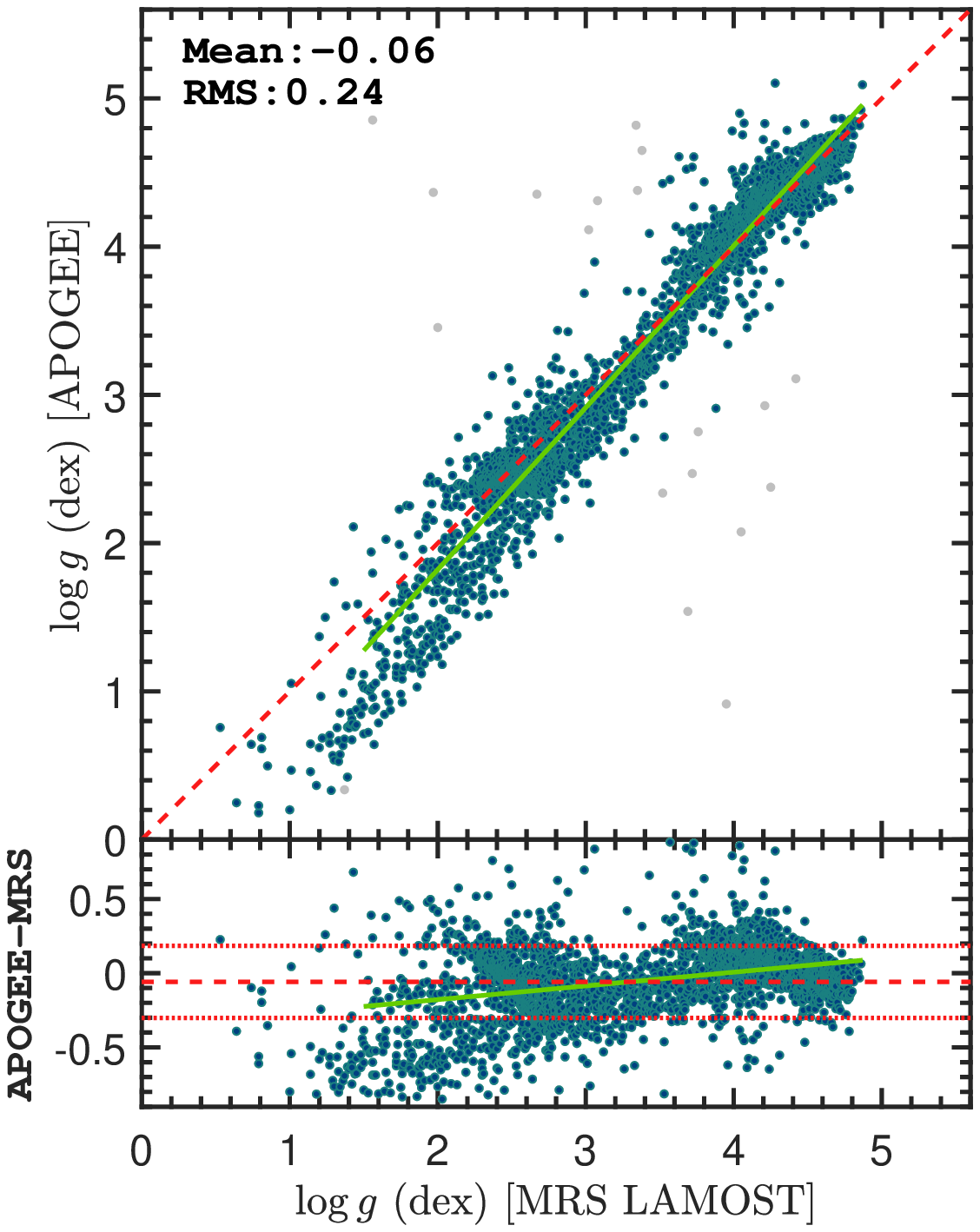}\includegraphics[width=6cm]{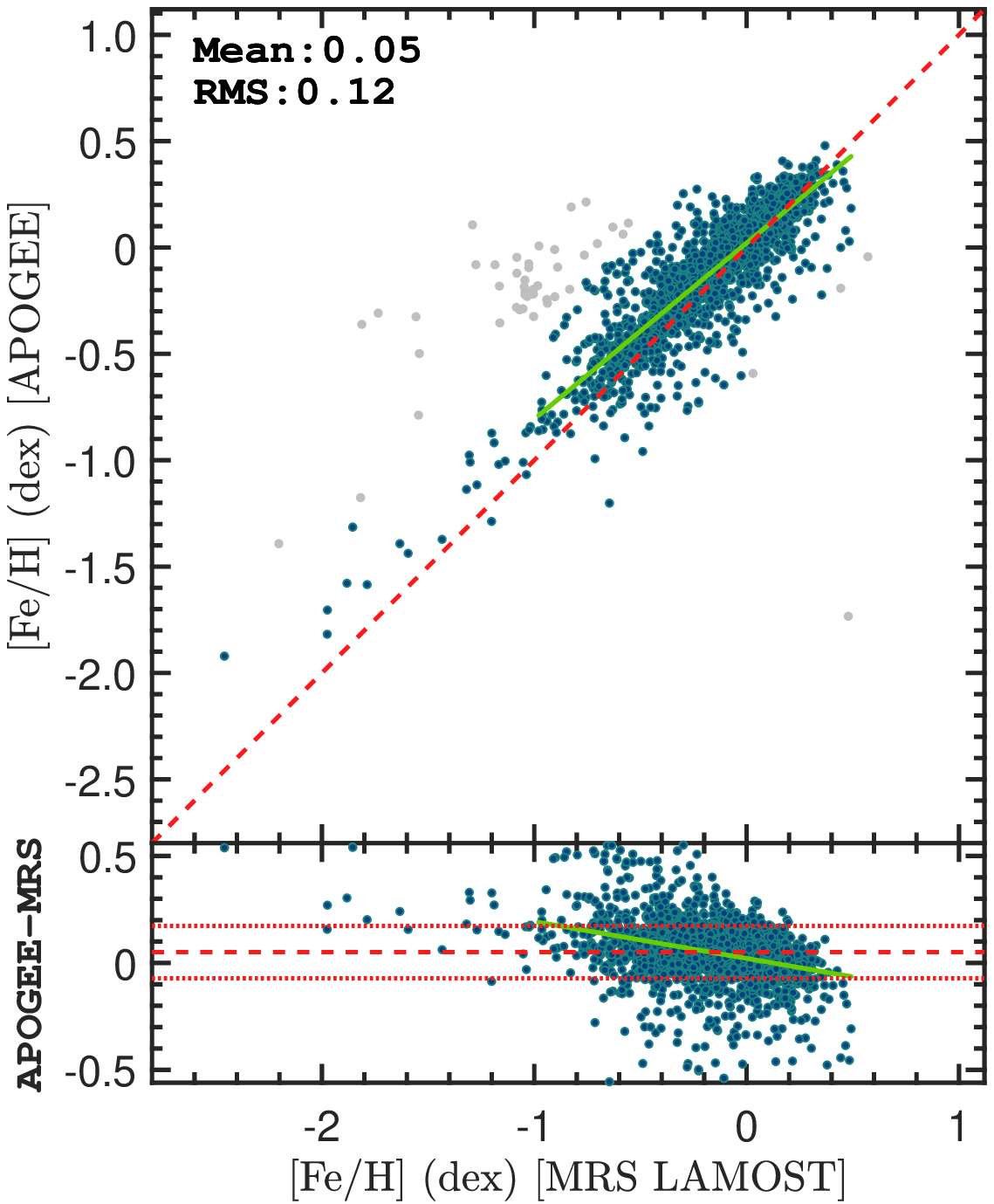}
\caption{Same as Figure\,\ref{lmrs} but for APOGEE and MRS LAMOST spectra with 1,514 stars in common. \label{ma}}
\end{figure*}

The Apache Point Observatory Galactic Evolution Experiment (APOGEE) was designed to solve the fundamental problem of galaxy formation through a systematic, homogeneous spectroscopic survey sampling all major populations of the Milky Way \citep{2017AJ....154...94M}. That program provides high-resolution, infrared spectra for  {$\sim$430,000 stars up to DR16 \citep{2020ApJS..249....3A}}, with pipeline-derived stellar atmospheric parameters and individual elemental abundances \citep{2018ApJS..235...42A}. The cross-identification between the LK-MRS and APOGEE surveys resulted in 2,749 common stars. Figure\,\ref{ma} shows comparisons of $T_\mathrm{eff}$, $\log g$ and [Fe/H] between those two catalogs where the latter two are the calibrated values. 
The majority of the objects located in the range [4000, 6500]\,K have $T_\mathrm{eff}$ values that are consistent with each other. However, over a wider range of effective temperatures the $T_\mathrm{eff}$ values of the two catalogs are related linearly as depicted by the regression line for $T_\mathrm{eff}<7000$\,K (considering hot stars are very few), but with a slope of 0.89 which is significantly different from unity. 
{The linear regression for $\log g$ shows a relation with a slope of 1.09 and an offset of -0.36 dex between the two catalogs down to a value of 1.5, without considering a few stars below that value.  However, below $\log g \sim 2.4$ dex, there is a clear bifurcation in the values.} A similar bifurcation, but less significant, is present in the comparison of $\log g$ from the California-$Kepler$ Survey and the Stellar Parameters Classification tool for $\log g$ values below $\sim$4.1\,dex (see mid-panel of Figure 16 in \citet{2017AJ....154..107P}). We note that the LASP pipeline gives a relatively large scatter when it is applied to giant stars with a low surface gravity \citep{lamost2015}. {In general, there is still a structure in the residuals if only the linear fitting applied, which needs polynomial fitting of the forth order to eliminate.} For the [Fe/H] parameter, most of the stars in common have metallicities in the range [-0.8, 0.4]\,dex, and so we applied the linear regression only to stars with [Fe/H] $>-1.0$. The linear regression line has a slope of 0.84 which is significantly different from unity in the same sense as the comparison with the LRS data.

\begin{figure}
\centering
%\epsscale{.80}
\includegraphics[width=8cm]{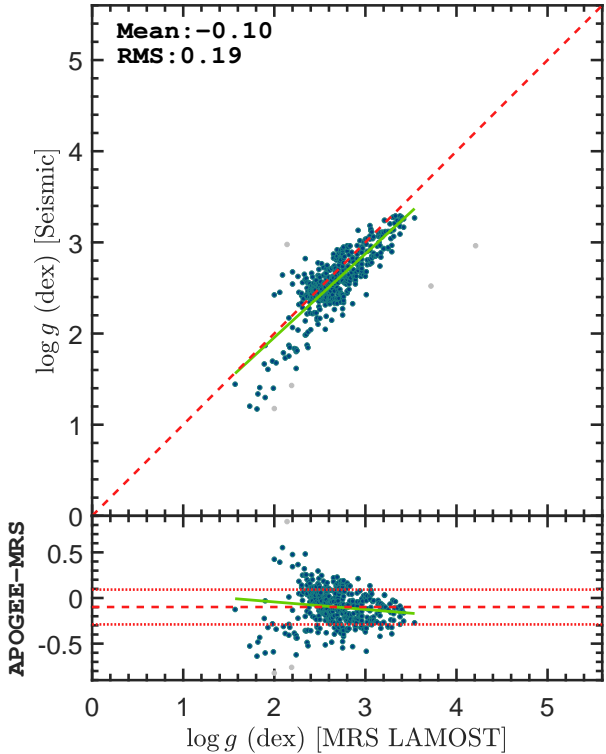}
\caption{Same as Figure\,\ref{lmrs} but for asteroseismic $\log g$ and MRS LAMOST spectra with 448 stars in common. \label{loggss}}
\end{figure}

{As the APOGEE $\log g$ of giants is calibrated with the asteroseismic results from {\sl Kepler}, we also show the direct comparison to the asteroseismic  $\log g$ in Figure\,\ref{loggss}. The cross-match results identify 448 common stars between the catalog of \citet{2018ApJS..239...32P} and our Table\,\ref{T5}. The linear regression shows a slope of 0.92 and an offset of 0.12 dex between the two catalogs.}

\subsection{GAIA}
% ------------------------------------------------------------------
\begin{figure*}
\centering
%\epsscale{.80}
\includegraphics[height=8.2cm]{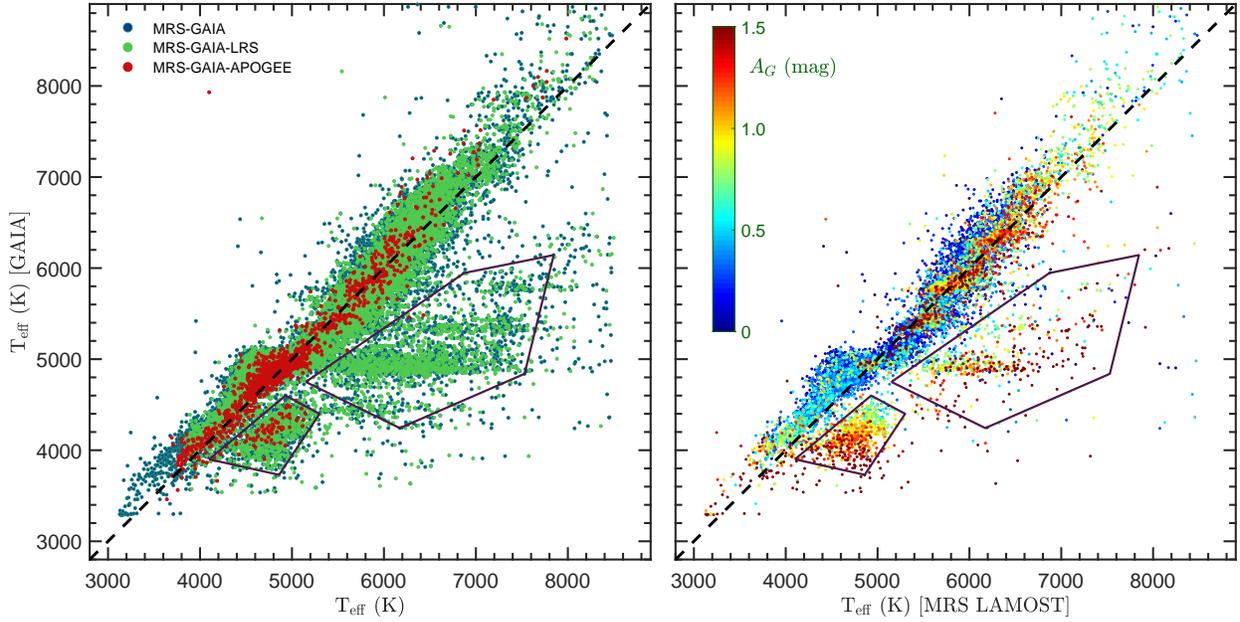}
\caption{{
Left panel: the comparison of the $T_\mathrm{eff}$ values for the 20,091 stars in common between the GAIA and LK-MRS survey (blue dots) as well as those in common with
the APOGEE (red dots) and LRS (green dots) surveys.  The abscissa plots the 
$T_\mathrm{eff}$ values from those three catalogs. Right panel: the comparison between the GAIA and LK-MRS $T_\mathrm{eff}$ values, but with extinction as an extra dimension (colors) for the 13,101 in-common targets with a value of $A_G$ from the GAIA catalog. In both panels the polygons outline the areas where the most discrepant outliers are found. Note that most of the outliers have large extinction values.  The dashed lines in both panels represent the bisector lines. 
}}
\label{mg}
\end{figure*}

GAIA is a space mission of the European Space Agency (ESA) {designed} to collect astrometry, photometry, RVs, and other astrophysical parameters for sources as faint as magnitude 21 in the GAIA $G$-band \citep{gaia2016a,gaia2018}. For $\sim161$ million sources with $G\leq17$, values for $T_\mathrm{eff}$ are calculated. These values range from 3000 to 10\,000\,K and have a typical accuracy of $\sim300$\,K \citep{2018A&A...616A...8A}. The cross-match of the LK-MRS and Gaia sources resulted in 20,091 objects in common, i.e. nearly the entire sample of LK-MRS objects. Figure\,\ref{mg} shows the comparison of the $T_\mathrm{eff}$ values in these two catalogs. The majority of the values are consistent with each other, as shown by the linear relation, but the scatter in the residuals is large. We find that a small fraction of objects show a significant difference in $T_\mathrm{eff}$. {The left panel shows the comparisons between the GAIA $T_\mathrm{eff}$ values and those from the three surveys LK-MRS (blue dots), LK-LRS (green dots) and APOGEE (red dots). It can be seen clearly that large discrepancies with the GAIA catalog can be found in all three of those catalogs. In particular, the majority of the discrepant points can be found in the two polygons drawn in that figure. This suggests that the discrepancies must be caused by the GAIA parameter pipeline since we do not observe similar discrepancies in the comparisons between the LK-MRS and LK-LRS catalogs (Figure\,\ref{lmrs}) and the LK-MRS and APOGEE catalogs (Figure\,\ref{ma}).}

{We note that} the GAIA parameter pipeline incorporates a machine learning algorithm to derive $T_\mathrm{eff}$ from the color indices $G_{BP}-G$ and $G-G_{RP}$ \citep{2018A&A...616A...8A}. The right panel {of Figure\,\ref{mg} } shows the result that the highly discrepant points generally have large line-of-sight extinction values with $A_G> 0.8$\,mag. {As one third of the stars in common between the GAIA and LK-MRS surveys do not have a GAIA $A_G$ extinction value, this prevents us from removing all of the high extinction stars when computing the bias and RMS values for the comparison between those two catalogs. However, if we restrict the comparison to only those stars with $A_G$ values (right panel of Figure\,\ref{mg}), we find a bias value of -8 K and an RMS of 303 K when stars with $A_G > 0.8$ are rejected.} \citet{2018A&A...616A...8A} claim that the {GAIA} $T_\mathrm{eff}$ values show a strong correlation with $A_G$. From the training result, they found that the RMS is 381\,K in $T_\mathrm{eff}$ when comparing the GAIA and LAMOST LRS values. Our MRS results are in good agreement with theirs once the more highly reddened outliers are removed, suggesting that lightly reddened GAIA $T_\mathrm{eff}$ values have an uncertainty of $\sim 300$\,K \citep{2018A&A...616A...8A}.

%% ------------------------------------------------------------------

\subsection{Radial velocity comparisons}
The LAMOST LRS, GAIA, and APOGEE surveys are used as external sources to estimate the accuracy of the RV determinations from the LK-MRS spectra. First, we removed {from our sample} the stars with only one RV measurement and targets for which the scatter in the individual RV measurements is larger than 0.5\,km\,s$^{-1}$ because those stars either have no information or have a high probability of being RV variables.

\begin{figure}
\centering
%\epsscale{.80}
\includegraphics[width=8.2cm]{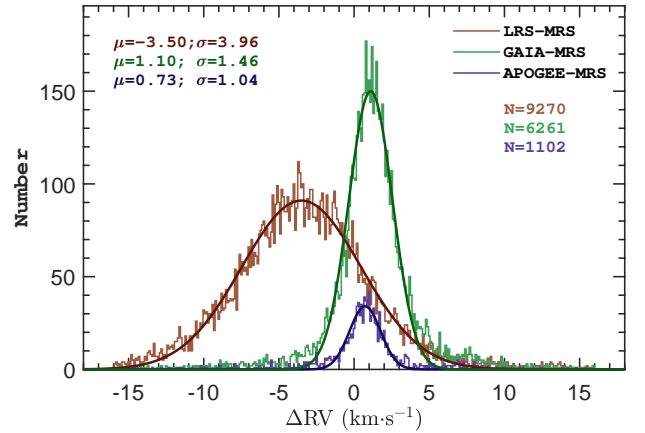}\\
\caption{{RV difference of LK-MRS to LAMOST LMR (brown histogram), GAIA (green histogram) and APOGEE (blue histogram).
Solid black lines represent the best Gaussian fittings whose centers and widths, $\mu$ and $\sigma$, are quoted in the legend.}}
\label{rvcom}
\end{figure}

{As it turns out, in a few cases, certain objects were observed at different epochs with an alternate wavelength calibration lamp.  Radial velocities derived from that lamp are systematically larger by 6.5 km s$^{-1}$ than those from the usual Th-Ar lamp (see Appendix\,\ref{AppRV}). In Table 3 that systematic correction has been applied, and that same correction will be applied to the spectra before they are released to the public.  With that correction applied, the} results of the comparisons with the LK-LRS, GAIA, and APOGEE catalogs are shown in Figure\,\ref{rvcom}.

We find that for the 9270 stars in common between the LAMOST LRS and MRS surveys the RV differences display a clear unimodal distribution.
This histogram is best fitted with a single Gaussian distribution with mean $\mu = -3.50$\,km\,s$^{-1}$ and standard deviation $\sigma = 3.96$\,km\,s$^{-1}$, which are the systematic offset and combined uncertainty of the two datasets. Indeed, the latter should be considered as the sum in quadrature of the typical uncertainties of the MRS and LRS data:
\begin{equation}
\sigma = \sqrt{\sigma_\mathrm{LRS}^2 + \sigma_\mathrm{MRS}^2}. \label{sig}
\end{equation}
However, given that the RV precision of LAMOST MRS spectra is much higher than that of the LAMOST LRS spectra, this $\sigma$-value allows an estimation of 3.8\,km\,s$^{-1}$ for the latter. This is very similar to the most recent estimation obtained by J. T. Wang, et al. 2020 (in prep.).

The comparison to GAIA is based on the 6261 objects with RV measurements in both catalogs and likewise results in a unimodal distribution. 
The histogram was fitted successfully with a single Gaussian distribution with $\sigma$-value ($\sigma \approx 1.46$\,km\,s$^{-1}$) and $\mu$-value ($\mu = 1.10$\,km\,s$^{-1}$), which gives the combined uncertainty and the offset between these two systems. Considering that $\sigma_\mathrm{MRS} \sim 1.0$\,km\,s$^{-1}$, the average precision of the Gaia RV values is estimated to be 1.1\,km\,s$^{-1}$. This is compatible with the results of \citet{2019A&A...622A.205K}, who gave a precision of 0.3\,km\,s$^{-1}$ for bright stars (4-8 mag) and 1.4-3.6\,km\,s$^{-1}$ for fainter stars ($\sim 12$\,mag) in the temperature range of 4000-6000\,K.

A relatively small sample of 1102 stars in common between APOGEE and LK-MRS is available for the RV comparison. Again, a clear unimodal distribution of RV difference is found with a slightly smaller standard deviation ($\sigma \sim 1.04$\,km\,s$^{-1}$) and  different mean value ($\mu_1 = 0.73$\,km\,s$^{-1}$). From Eq.\,(\ref{sig}), we can conclude that the precision of the APOGEE RV is better than that of LK-MRS where the latter contributes the large uncertainty. This result is in full agreement with Figure 24 of \citet{2015AJ....150..173N}.

\section{Science prospective}
With just one year of LAMOST observations, the LK-MRS project has acquired about 280,000 blue and 370,000 red spectra of 28,000 objects with ${\rm S/N}>10$, 55\% of which have high-quality photometry from {\sl Kepler/K}2. This is the first time, to our knowledge, that a large spectroscopic survey has been dedicated to monitor more than 50,000 stars with multiple visits (see Table\,\ref{T1}). The derived parameters have an internal precision of 100\,K, 0.15\,dex and 0.09 dex for $T_\mathrm{eff}$, $\log g$ and [Fe/H] at ${\rm S/N}\sim 10$, respectively. These have been estimated based on multiple measurements of the same star. However, the determined parameters, which are derived from a pipeline, may be affected by systematic errors, as shown by comparisons with other surveys, which display, in some cases, significant discrepancies. This can be particularly relevant for metallicity, especially in the domain of metal-poor stars for which there are only a few measurable lines for a given element of interest within the wavelength range covered by the MRS spectra. For those who pay attention to certain special targets, {we remark that the stellar parameters must be taken with care, especially for very metal-poor stars.} We encourage other groups to independently calculate atmospheric parameters from the LK-MRS spectra with their own pipelines. Within the LAMOST LRS survey, observations of plates covering the {\sl Kepler/K}2 fields have been carried out in order to gather a homogeneous collection of spectra for as many targets with high-quality photometry as possible (Z18b; Wang et al., in prep.). The LK-MRS survey, however, is dedicated to the monitoring of 20 plates with up to 3000 objects each at multiple epochs over a time span of five years, for the purpose of investigating the variations of physical parameters such as RV, the width of the line profiles, $\log g$, etc. This {observing strategy is producing data very suitable for various research fields, such as the study of multiple stellar systems, stellar activity, and pulsating stars.}

\subsection{Multiple systems}
Multiple systems (binaries, triple stars, etc.) are important for 
testing and refining stellar evolution theory, from star formation to the final stages of evolution \citep[see, e.g.,][]{2004MNRAS.350.1301H,2013ARA&A..51..269D}. 
The general physical properties {of multiple systems, such as}
the distribution of orbital periods and eccentricities, can be obtained through the monitoring of RV variations for a large sample of stars \citep[see, e.g.,][]{1991A&A...248..485D,2010ApJS..190....1R}. 
{Wide area photometry in rich stellar fields is} an effective method for discovering binaries with periods {ranging from minutes to years} \citep{2011AJ....141...83P} despite the fact that the detection probability for eclipsing events is low because of the requirement for an inclination angle $i$ close to 90 degrees. 
Even with high-quality photometry from space, only a few percent of the observed stars turn out to be eclipsing binaries (e.g. a few thousands of the $\sim 200,000$ stars observed by {\sl Kepler}). {Spectroscopy, however, can yield much higher detection rates.} A tentative simulation based on one plate from the LK-MRS survey suggests that the percentage of binary systems is actually higher than 10\% (J. X. Wang et al., in prep). Indeed, roughly 200 out of 1900 stars revealed regular RV variations with a amplitude higher than 3$\sigma_\mathrm{RV}$ (defined by  Eq.\ref{eqpower}). This test predicts that on the order of 5,000 binary systems will be detectable from the LK-MRS survey, {making possible the construction of a large, unbiased sample of binaries.} As LK-MRS aims to have $\sim55\%$ {of the stars in its sample} in common with {\sl Kepler/K}2 (column 'KO' in Table\,\ref{T1}), {it will be possible to provide} a direct estimate of the ratio of eclipsing binaries to RV binaries. This ratio could be used to deduce the distribution of $i$ for short-period binaries, as eclipsing events require that the primary and secondary components must occult each other. {A limitation is that the LASP pipeline derived parameters and RVs treating the spectra belong to one single stars. Independent works are encouraged to identify multiple systems and provide their own parameters and RVs.}

The RV variations, detected by the LK-MRS survey, will reveal a large sample of orbital companions with diverse populations, fertilising the field of stellar evolution theory. 
RV variations with a large peak-to-peak amplitude can be helpful 
for discovering unseen massive compact companions such as a neutron star or a black hole \citep[see, e.g.,][]{2019ApJ...872L..20G}. 
With the help of LAMOST LRS spectra, \citet{Liunature2019} recently announced the discovery of a massive black hole companion orbiting a B-type star, LB-1. 
This result supports the contention that LAMOST is an ideal instrument for hunting RV variables. 
We note that the MRS spectra have a RV precision better than 1\,km\,s$^{-1}$, as illustrated in Figure\,\ref{err}, which is about $3\sim4$ times better than that of the LRS spectra. 
The MRS spectra with ${\rm S/N}>60$ have a precision better than $0.5$\,km\,s$^{-1}$, which offers the opportunity of detecting  companion masses as low as those of brown dwarfs. 
Therefore, high-quality time series in the LK-MRS survey have the potential to provide a few brown dwarf companions {with masses just below the minimum stellar mass limit.}
However, we may encounter the notable dearth of brown dwarf companions in short period orbits, a.k.a. the brown dwarf desert \citep{2006ApJ...640.1051G,2019A&A...631A.125K}, which may provide clues to the understanding of tidal interaction/dissipation and magnetic braking between the substellar companion and the host star \citep{2018MNRAS.481.4077S,2014EAS....65..327G}. 

\subsection{Pulsating stars}
Time-series of spectroscopic observations are particularly helpful for mode identification of pulsating stars through methods based on line-profile variations, such as the moment method \citep{1996A&A...314..115A} and the Fourier-parameter-fit method \citep{2006A&A...455..227Z}. But those techniques require high-resolution spectra to resolve the line-profile variations. The resolution of the LAMOST MRS ($R\sim7500$) is insufficient for this purpose. This is the main drawback of most large spectroscopic surveys employing multiple fibers. However, the LAMOST MRS spectra will provide multiple measurements of atmospheric parameters of pulsating stars, allowing a precise determination of their location in the Kiel-diagram, and thus the determination of the empirical borders of the instability strips. Hence, these observations are providing mandatory ingredients for the seismic modeling of such stars with pulsation codes \citep[see, e.g.,][]{2018Natur.554...73G,charpinet2019}. 
Another benefit of time-series spectra is that atmospheric parameters of a pulsating star can be calculated at different epochs, thus revealing changes in its location in the HR-diagram.
In particular, large amplitude pulsators with monoperiodic brightness modulations,
such as RR\,Lyrae stars and Cepheids, might exhibit periodic changes of $T_\mathrm{eff}$ and $\log g$.

With {\sl Kepler/K}2 photometry, we can measure the frequencies of modes of oscillation with a very high precision, down to a few nHz \citep[see, e.g.,][]{zong2018a}. {On the other hand, the accuracy of the determination of the intrinsic (bolometric) photometric amplitudes of pulsating stars is compromised by factors such as uncertainties in the bandpass efficiencies, as well as uncertainties in the Galactic extinction. This compromises our ability to obtain good agreement between model fits and the observed bolometric amplitude. } In addition, the current linear seismic models 
are not able to determine the amplitude of an oscillation mode, {as that calculation requires the inclusion of} higher order nonlinear perturbation terms. {Conversely, the periodic RV variations which arise from the pulsation of the star, especially those associated with radial modes, can be well described by current stellar pulsation models} \citep[see, e.g.,][]{2008AcA....58..193S}.
Therefore, precise pulsation periods are first measured from {\sl Kepler/K}2 photometry {in order to derive an accurate ephemeris for the associated
RV variations.}
In turn, the optimal seismic models for (large-amplitude) pulsating stars can be constructed with the help of these RV measurements. 
We note that, unlike the intrinsic amplitude of a pulsation mode, the RV associated with a particular pulsation mode is an intrinsic physical quantity that does not suffer from external contamination such as Galactic extinction.

Multiple systems may contain one or more pulsating components whose physical quantities, such as mass, can be obtained through different independent methods, such as the orbital solution and/or asteroseismology \citep[see, e.g.,][]{2008A&A...489..377C}. 
Recently, \citet{2018MNRAS.474.4322M} discovered 341 binary systems 
out of 2,000+ pulsating A/F stars observed with {\sl Kepler} %light curves 
from phase modulation of their pulsations. 
The binaries in their samples have orbital periods ranging from a few months to years, suitable for spectroscopic confirmation with data from a survey such as the LK-MRS survey.
Their method can be applied to {\sl Kepler} photometry of other types of pulsating stars, provided that their pulsation frequencies are stable \citep{2018haex.bookE...6H,zong2018a}.

\subsection{Stellar activity}

Stellar activity, a term which encompasses a range of phenomena produced by dynamo action in stellar interiors, is related to magnetic fields and, in turn, to stellar rotation, differential rotation, and sub-photospheric convection. The magnetic activity manifests itself through several phenomena, such as radio and/or X-ray coronal emission, UV and optical chromospheric emission lines,  sudden brightness variations (in the continuum or in spectral lines) known as flares, and rotational modulation of brightness produced by cool spots.
The long uninterrupted monitoring of thousands of stars by the {\sl Kepler/K}2 mission offers a unique opportunity to investigate stellar activity, through optical photometry, for a large sample of stars with different spectral types, masses and ages \citep[see, e.g.,][]{2019ApJS..243...28L,2019ApJS..241...29Y,2019ApJ...871..241D}. {Roughly $55\%$ of the {\sl Kepler/K}2 sources observed with LAMOST will be provided with time-domain spectra at about 60 different epochs by the LK-MRS project. Useful activity indicators in the optical domain include the equivalent width of chromospheric lines such as Balmer H$\alpha$, Ca\,{\small $\mathrm{\RNum{2}}$}\,IRT (in the near infrared) and {Ca\,{\small $\mathrm{\RNum{2}}$}}\,$H$ and $K$ lines at about 393\,nm \citep[see, e.g.,][]{2008AJ....135..785W,2016A&A...594A..39F}. Medium-resolution spectra, provided by LK-MRS, fully cover the H$\alpha$ emission line in the red-arm spectra. This enables the investigation of the variability of H$\alpha$ emission for active stars at different epochs, arising from rotational modulation of chromospheric plages \citep[see, e.g.,][and reference therein]{Frasca2000,Frasca2008} and/or by flares \citep[see, e.g.,][]{CatalanoFrasca1994,Foing1994}. The numerous stars exhibiting flare events from photometry, can also be thoroughly investigated for the flare properties by combining the photometric information with that gained from the MRS spectra. {More specifically, the flare frequency and energy budget can be studied as a function of the average activity level derived from the H$\alpha$ chromospheric emission.}}

Magnetic activity shows a significant correlation with the stellar rotation period, $P$, or angular velocity, $\Omega=2\pi/P$. As mentioned above, the projected equatorial velocity $v\sin i$ can be measured from MRS spectra of relatively rapid rotators ($v\sin i\geq$\,15\,km\,s$^{-1}$). Although this parameter is affected by the stellar radius and the projection factor $\sin i$, for a large sample of stars it may be used to investigate the dependence of magnetic activity level on stellar rotation, at least in the middle/high activity regime. However, for several stars in the LK-MRS project, we know the rotation period from the high-precision {\sl Kepler/K}2 photometry and the radius can be evaluated thanks to the MRS atmospheric parameters (basically $T_{\rm eff}$) and the {\it Gaia} parallax. Therefore, for those stars, the inclination of the rotation axis can be determined.

\subsection{Other interesting objects}
Similar to the LAMOST LRS survey, the LASP pipeline for the MRS spectra provides parameters only for late-A, F, G, and K stars. Apart from those objects, there is a relatively small sample of stars of particular importance, such as hot and highly evolved stars including the O/B type main-sequence stars, the white dwarfs and the hot subdwarf stars (sdOB). We note that more than $50\%$ of the stars of the latter group reside in short-period binary systems. However, this claim is based on a statistical survey of only a few dozens of sdB stars \citep[see, e.g.,][]{2001MNRAS.326.1391M}. The time resolved MRS spectroscopy from LAMOST will not only shed new light on this field but will also allow the determination of the chemical abundance of specific elements for those stars based on non-local thermodynamic equilibrium atmospheric models \citep{2009ARA&A..47..211H,2019ApJ...881..135L,2019ApJ...881....7L}. White dwarfs, as the graveyard of most low mass stars, are an astrophysical laboratory to test physics under extreme conditions, such as dark matter-electron interactions \citep{2018PhRvD..98j3023N}. 
Several independent studies based on the LRS survey of LAMOST, including those from the LK-project, have been concentrating on the characterization and determination of atmospheric parameters for such stars \citep[see, e.g.,][]{2015MNRAS.454.2787G}. The results obtained for these types of stars could be improved by using the spectra of the MRS survey of LAMOST.

\section{Summary}
The LAMOST-{\sl Kepler/K}2 MRS survey (LK-MRS), initiated in 2018, aims at collecting medium-resolution ($R\sim7,500$) spectra for more than 50,000 stars. It is one of the four parallel projects dedicated to collect time series of spectra and will observe all these stars about 60 times over a period of 5 years (from 2018 September to 2023 June). In accordance with the allocated time, we selected 20 footprints distributed across the {\sl Kepler} prime field and six {\sl K}2 campaigns in the northern hemisphere, with each plate containing typically $\sim2,000$ to $\sim 3,000$ stars. The input catalog of the LK-MRS survey includes about $94\%$ and $53\%$ stars in common with the {\sl Kepler/K}2 input catalog and the list of stars for which space-based photometry has been collected, respectively. Almost all stars of the project have a GAIA $G$-band magnitude brighter than 15, as shown in Figure\,\ref{fig1}.

After one year, a total of 223 exposures have already been gathered for 13 different plates during 40 individual nights. Each plate has been visited between 4 and 46 times (see Figures\,\ref{expo}). We have collected about 280,000 and 369,000 high-quality spectra ($\mathrm{S/N}>10$) in blue and red wavelength ranges, respectively. For the objects with a spectral type ranging from late-A to K, atmospheric parameters and RVs were derived. This could be done successfully for about 259,000 blue MRS spectra of 21,053 targets. The distribution of weighted average values for these parameters is shown in Figure\,\ref{swan} (Kiel diagram) and \ref{fours} (histograms). Their values are listed in Table\,\ref{T5}. The internal uncertainties for $T_\mathrm{eff}$, $\log g$, [Fe/H], and RV are evaluated through the  measurements of the same objects at multiple epochs. They are estimated to be 100\,K, 0.15\,dex, 0.09\,dex, and 1.00\,km\,s$^{-1}$ when derived from MRS spectra with $\mathrm{S/N}=10$, respectively. These uncertainties decrease for increasing values of S/N, but they stabilize for $\mathrm{S/N}>100$ (see Table\,\ref{abc}). These precisions reach the objectives of the design of the LAMOST MRS survey {(C. Liu, private communication).} 

We compared our parameters and RVs with those of the LAMOST LRS \citep{lamost2015}, APOGEE \citep{2017AJ....154...94M}, and GAIA \citep{gaia2018} surveys to check the quality of our results. There are, 14,997, 1,514 and 20,091 stars in common, respectively, corresponding to a fraction of $\sim70\%$, $\sim7.2\%$ and $\sim95\%$ of the LK-MRS survey. In general, the LAMOST MRS parameters are consistent with the LAMOST LRS measurements (Figure\,\ref{lmrs}), likewise for the comparison with APOGEE, but the scatter increases as $\log g$ decreases (Figure\,\ref{ma}). A large discrepancy was found in the $T_\mathrm{eff}$ comparison with GAIA. This is mainly due to the fact that the $T_\mathrm{eff}$ values from GAIA are derived from a color-index (Figure\,\ref{mg}). The comparisons of RVs show unimodal Gaussian distributions where the offset values of LK-MRS to LAMOST LRS, LK-MRS to GAIA and LK-MRS to APOGEE are -3.50, 1.10 and 0.73\,km\,s$^{-1}$, respectively (Figure\,\ref{rvcom}).

The LK-MRS survey is the first project dedicated to obtaining time series of spectra by using the LAMOST MRS spectrographs, pointing towards the {\sl Kepler/K}2 fields. These spectra will be very important for many scientific goals, including the discovery of new binaries, the study of oscillation dynamics for large-amplitude pulsators and the investigation of the variability of stellar activity. From a preliminary simulation, we expect that at least 5,000 binaries will be detected by the end of the survey phase in 2023, solely through the technique of RV variations. We encourage other groups to develop their own pipelines for analysing the LK-MRS spectra. All the spectra discussed in this paper will be available after 2021 September.
%------------------------------------------------------------

\acknowledgments
We acknowledge support from the Beijing Natural Science Foundation (No. 1194023) and the National Natural Science Foundation of China (NSFC) through grants 11673003, 11833002 and 11903005. The Guoshoujing Telescope (the Large Sky Area Multi-object Fiber Spectroscopic Telescope LAMOST) is a National Major Scientific Project built by the Chinese Academy of Sciences. Funding for the project has been provided by the National Development and Reform Commission. LAMOST is operated and managed by the National Astronomical Observatories, Chinese Academy of Sciences. WKZ holds the LAMOST fellowship as a Youth Researcher which is supported by the Special Funding for Advanced Users, budgeted and administrated by the Center for Astronomical Mega-Science, Chinese Academy of Sciences (CAMS). WKZ and JNF acknowledge the support from the Cultivation Project for LAMOST Scientific Payoff and Research Achievement of CAMS-CAS. This paper is dedicated to the 60th anniversary of the Department of Astronomy of Beijing Normal University, the 2nd one in the modern astronomy history of China. AF and GC acknowledge financial support from INAF. JM-\.Z acknowledges the Wroclaw Centre for Networking and Supercomputing grant no.224. The work presented in this paper is supported by the project "LAMOST Observations in the {\it Kepler} field" (LOK), approved by the Belgian Federal Science Policy Office (BELSPO, Govt. of Belgium; BL/33/FWI20). H.-L.Y. acknowledges support from NSFC (No. 11973052) and the Youth Innovation Promotion Association, CAS. This research is supported by the Astronomical Big Data Joint Research Center, co-founded by the National Astronomical Observatories, Chinese Academy of Sciences and Alibaba Cloud.

\software{LASP \citep[v2.9.7;][]{2011A&A...525A..71W,lamost2015}}

\bibliographystyle{aasjournal}

\appendix

\section{Radial velocity correction} \label{AppRV}
\begin{figure*}
\centering
%\epsscale{.80}
\includegraphics[width=16cm]{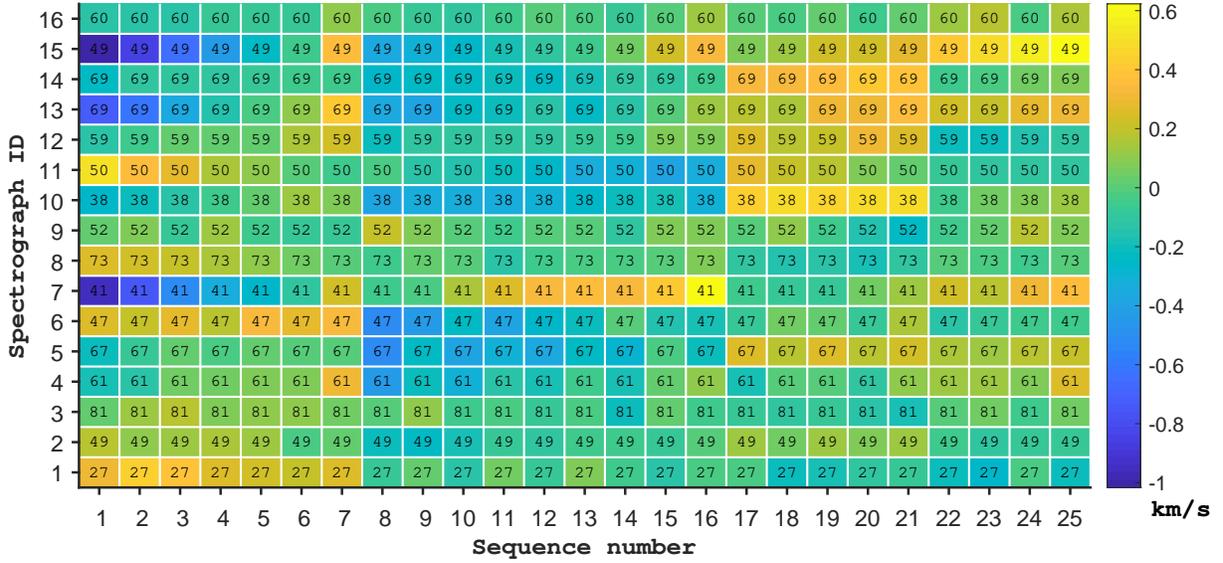}\\
\caption{
Example of distribution of the systematic offsets for the ``constant common'' stars as a 2D-function of the spectrograph ID and observed epochs for the test plate ``K1a1''. The marked numbers indicate how many common stars were used to calculate the offset values in km\,s$^{-1}$, which are represented by the colors. {The offsets are typically {on the order of} a few hundreds\,m/s.}
}
\label{a2}
\end{figure*}

L19 noted that small, systematic offsets exist in RV measurements based on LAMOST MRS spectra, resulting from slight zero-point variations of the calibrated wavelength.
They proposed a correction procedure to remove these offsets that is based on a specific characteristic of the observation strategy: in TD mode, a star on a plate is always observed with the same fiber.
The offset values are derived from multiple independent measurements on a large sample of stars. Considering the RV precision of MRS spectra, it is safe to assume that more than half of the objects in each plate are ``constant'' RV stars. The deviations of RV will be significantly reduced if the ``existing'' offsets are correctly removed. We first used the data set of L19 to test the precision of the offsets by doing three iterations of the procedure. The offset values from the first iteration are shown in Figure\,\ref{a2} where we can clearly see that the offset is a 2D function relative to the spectrograph ID and time (sequence number). 
A second iteration does not change the standard deviations of the shifted RVs significantly. 
%any more. 
We therefore conclude that the offsets are effectively removed after one round of corrections. From Figure\,\ref{a2}, we can see that the offsets are of the order of a few hundreds ms$^{-1}$. The offsets will become larger when the observations span a relatively long period, such as a few months. Finally, the systematic offsets were removed for all observed plates of the LK-MRS survey, resulting in the RVs listed in Table\,\ref{T5}.
We note that there are a few plates whose spectra were, {{on occasion, calibrated with an alternate arc lamp during the test phase and the beginning of the} 
2018 season. The RV difference 
%for those two arcs has been identified with the value of 6.5\,km\,s$^{-1}$. The plates, occasionally observed with two different lamp arcs, will be firstly shifted to the value of the Th-Ar lamps by adding 6.5\,km\,s$^{-1}$ before data public.
{resulting from the use of that alternate arc lamp has been determined to be 6.5\,km\,s$^{-1}$. Therefore, a radial velocity shift of 6.5\,km\,s$^{-1}$ will be applied to those spectra before the data are released to the public.}
}
%leading to two distinct peaks in distributions shown in Figure\,\ref{rvcom}. The use of two different arc lamps only affect the zero-points when combining the LK-MRS RVs with those from other surveys. It will not harm the results of the analysis of RV variations for single star if only RVs from the LK-MRS survey are used.

% ----------------------------------------------------
\section{Atmospheric parameter correction} \label{AppAP}
\begin{figure*}
\centering
%\epsscale{.80}
\includegraphics[width=5.75cm]{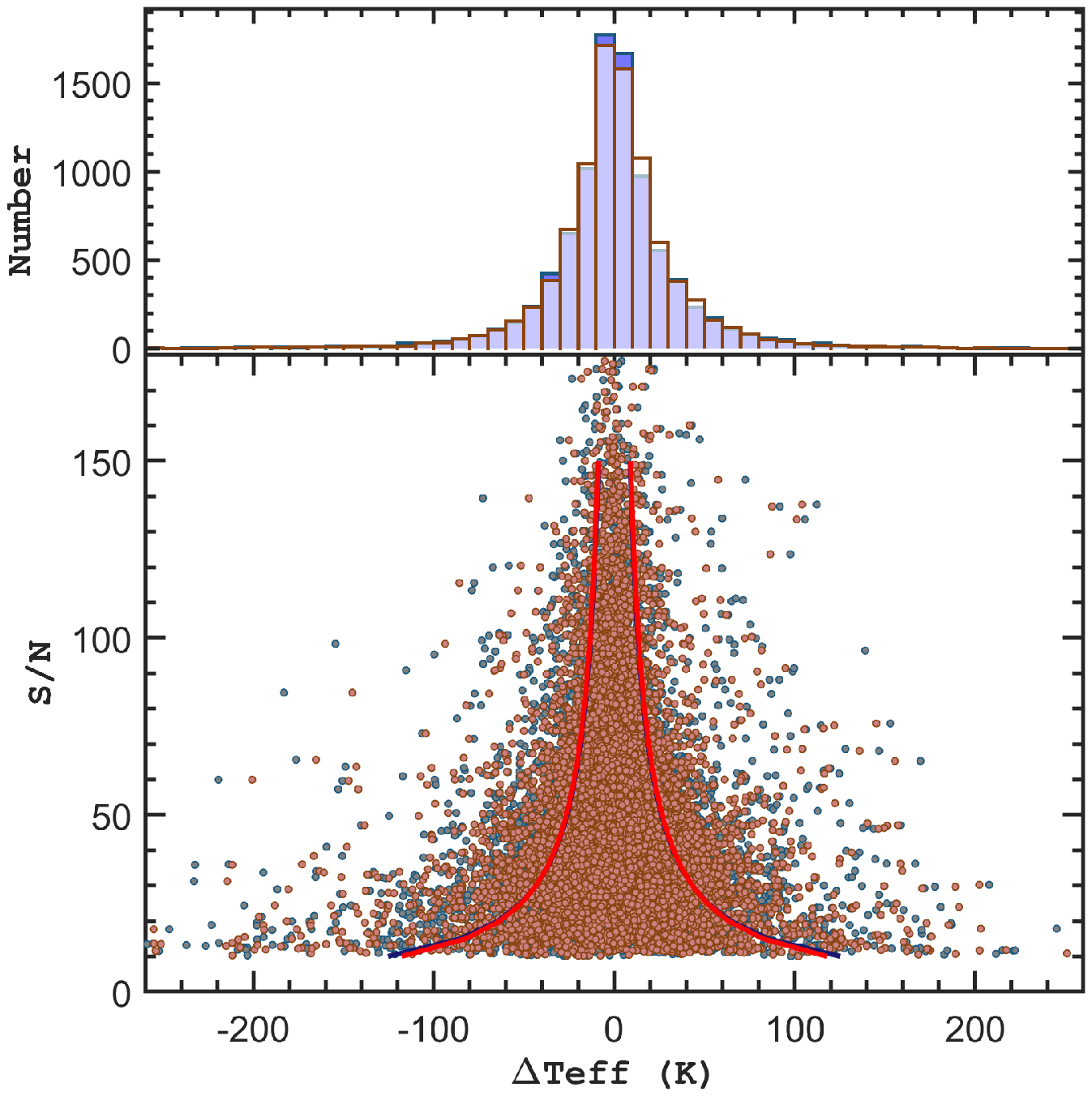}
\includegraphics[width=5.75cm]{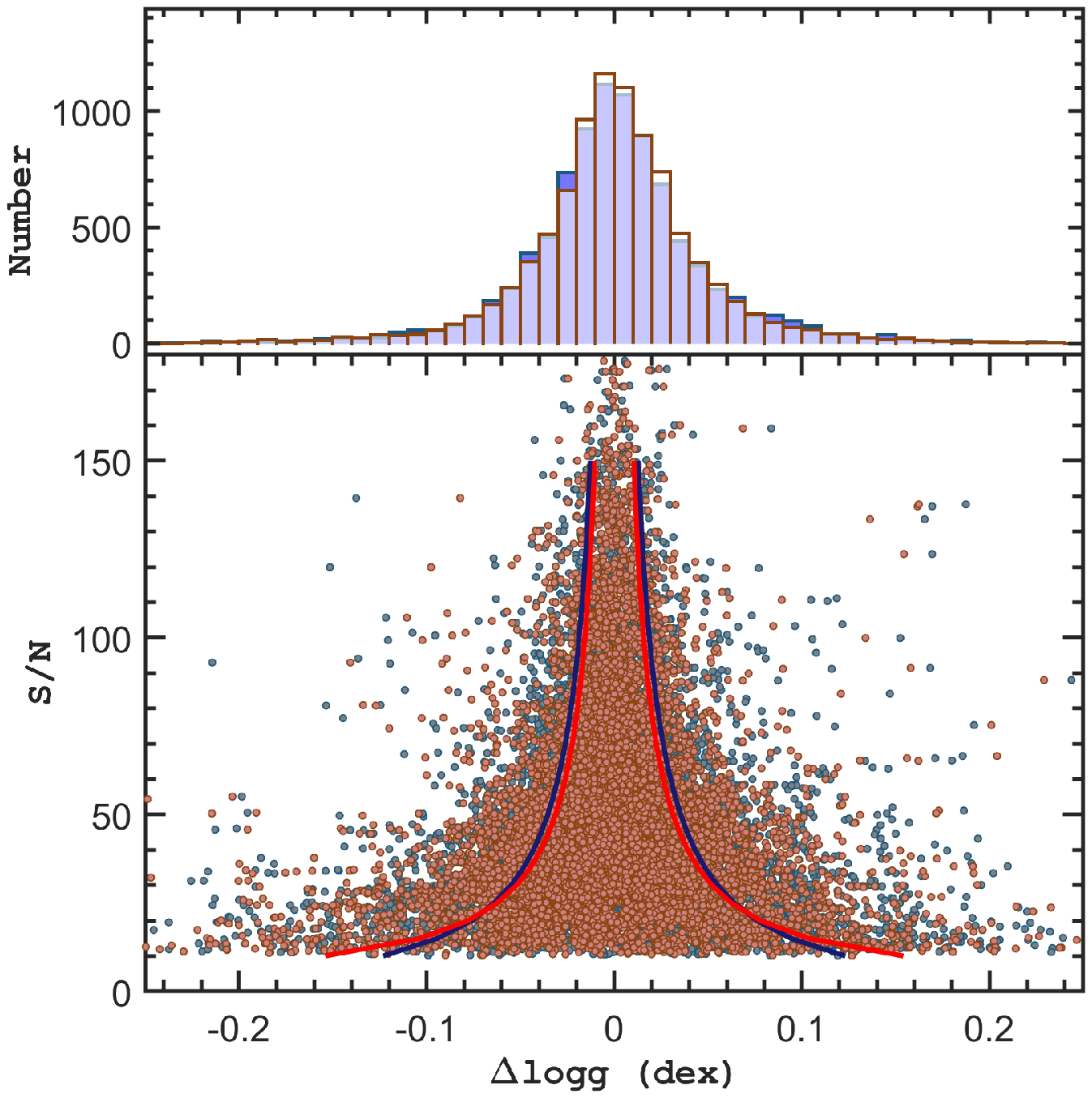}
\includegraphics[width=5.75cm]{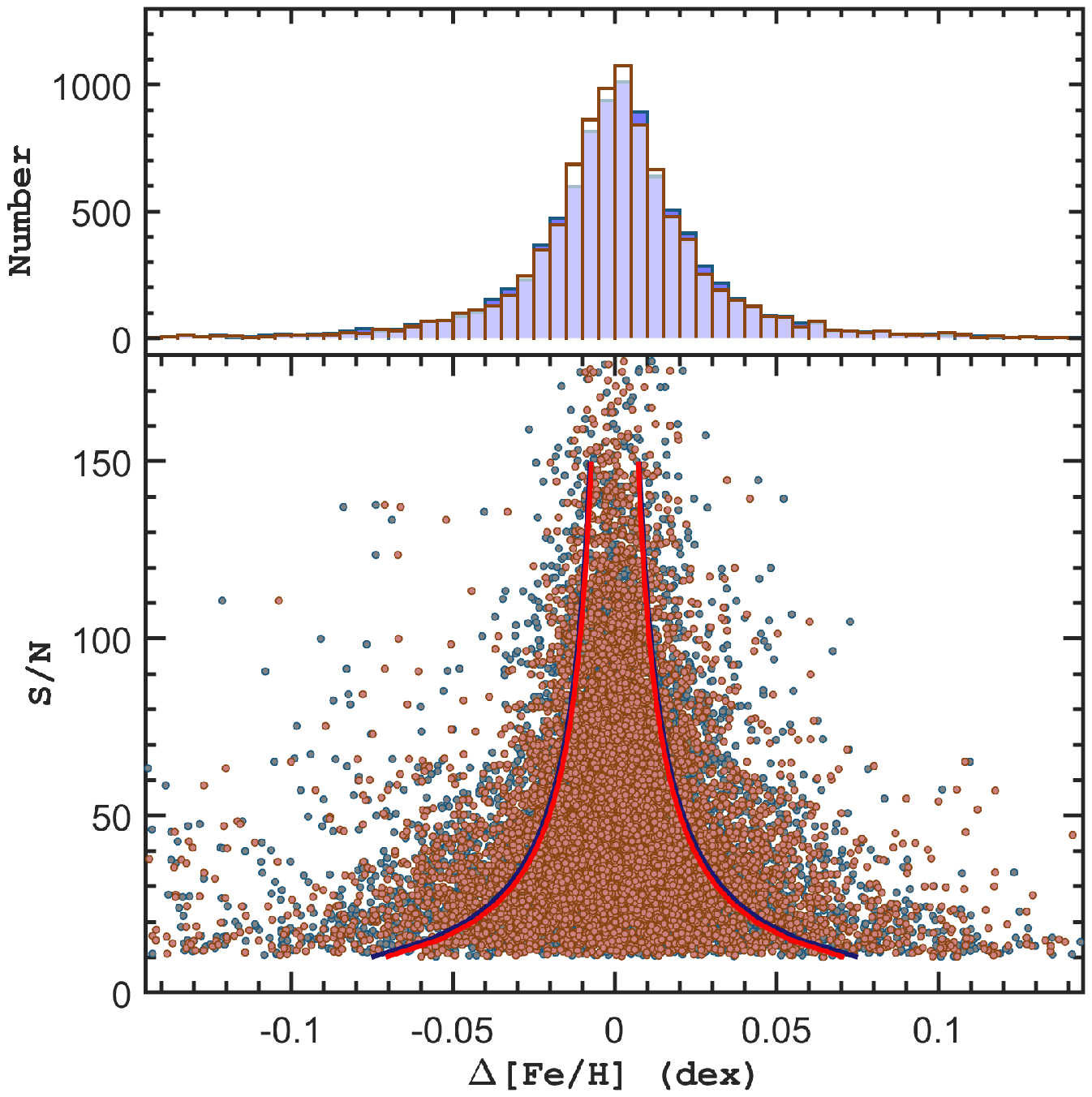}
\caption{
Distribution of the shifted atmospheric parameters $T_{\mathrm{eff}}$ (left), $\log g$ (middle), and [Fe/H] (right) derived from the plate TD085754N225914K01 (or K2d4), as a test to compare the results with (brown) or without (blue) correction of the systematic offsets. 
The solid curves represent the optimal fits as a function of S/N, where red and blue refers to the corrected and uncorrected parameters, respectively.
See details in the text.
}
\label{a1}
\end{figure*}
% ++++++++++++++++++++++++++++++++++++++++++++++
The correction of the small systematic offsets significantly improves the precision of RVs derived from the LK-MRS spectra. This section is dedicated to testing whether a similar correction process is suitable and necessary for the atmospheric parameters derived from the same spectra. 
This test was performed on the data obtained for plate TD085754N225914K01 (or K2d4). 
The observations of this plate are summarized in Table\,\ref{T2log}. 
The first step is to set up conditions for so-called ``common constant'' (CC) stars in each parameter. 
We only used the 24 archived spectra for this plate observed on 4 nights (5 exposures of 2018/11/25; 6 exposures of 2019/02/13; 6 exposures of 2019/02/21; 7 exposures of 2019/02/25). 
The data set over more days may reduce the number of common stars, i.e. the stars for which the atmospheric parameters can be derived from all the qualified spectra. 
We use somewhat arbitrary criteria for the atmospheric parameters to be considered as constant, i.e., $\Delta{T_\mathrm{eff}}<100$\,K, $\Delta\log g < 0.15$\,dex, and $\Delta \mathrm{[Fe/H]}<0.1$\,dex, respectively. 
It results in 379, 380, and 383 CC stars for each of the 24 visits for $T_\mathrm{eff}$, $\log g$, and [Fe/H], respectively. 
The correction procedures are similar to the RV corrections introduced by L19 (cf. Appendix\,\ref{AppRV}). 
The atmospheric parameters of the CC stars are divided into different groups by their sequence and spectrograph ID. 
Then a weighted vector of correction values can be calculated. 
We refer the interested reader to Section~3.4 of L19 for more details about this procedure.

Figure\,\ref{a1} shows the distribution of the atmospheric parameters shifted to their weight averaged values before and after correction. 
The differences of these parameters before and after correction are very small, which agrees with the fact that the correcting vector contains very small values. 
However, the correction induces small changes in the histograms for each type of parameter. 
The optimal fit of each parameter is built on the dependence of the standard deviations on S/N. 
The fit roughly gives the precision of the measurements as a function of their S/N. 
We clearly see that the correction leads to only a negligible improvement of the precision for $T_\mathrm{eff}$ and [Fe/H], as suggested by the observation that the fits almost overlap with each other (Figure\,\ref{a1}, left and right panels). 
For $\log g$, the precision improved at high S/N while it worsened at low S/N (Figure\,\ref{a1}, middle panels). 
We note that these differences are still very small: for instance, $\sim0.003$\,dex at ${\rm S/N}\sim100$, a value that is much smaller than the current measuring precision in $\log g$. 
We therefore conclude that the corrections for systematic offsets for atmospheric parameters derived from LK-MRS spectra are not significant. Hence, they were not applied.

\end{document}